\documentclass[pss]{wiley2sp} 
\usepackage{amsmath}
\usepackage{bm}              
\usepackage{w-greek}         

\begin{document}

\title{Spin pumping through quantum dots}

\titlerunning{Spin pumping through quantum dots}

\author{%
  Stephan Rojek\textsuperscript{\textsf{\bfseries 1}},
  Michele Governale\textsuperscript{\textsf{\bfseries 2}},
  J\"urgen K\"onig\textsuperscript{\Ast,\textsf{\bfseries 1}}}

\authorrunning{S. Rojek, J. K\"onig, and M. Governale}

\mail{e-mail
  \textsf{koenig@thp.uni-due.de}, Phone:
  +49-203-379-3329, Fax: +49-203-379-3665}

\institute{%
  \textsuperscript{1}\,Theoretische Physik, Universit\"at Duisburg-Essen and CENIDE, 47048 Duisburg, Germany\\
  \textsuperscript{2}\,School of Chemical and Physical Sciences and MacDiarmid Institute for Advanced Materials
and Nanotechnology, Victoria University of Wellington, PO Box 600, Wellington 6140, New Zealand}

\received{XXXX, revised XXXX, accepted XXXX} 
\published{XXXX} 

\keywords{quantum dots, tunneling, adiabatic pumping, spin currents, diagrammatic transport theory.}

\abstract{%
%
%
%
\abstcol{%
We propose schemes for generating spin currents into a semiconductor by adiabatic or non-adiabatic pumping of electrons through interacting quantum dots. 
The appeal of such schemes lies in the possibility to tune the pumping characteristics via gate voltages that control the properties of the quantum dot.
The calculations are based on a systematic perturbation expansion in the tunnel-coupling strength and the pumping frequency, expressed within a diagrammatic real-time technique.
Special focus is put on the possibility of pure spin pumping, i.e., of pumping spin currents without charge currents.
  }}
%
%
\titlefigure[width=8.1cm]{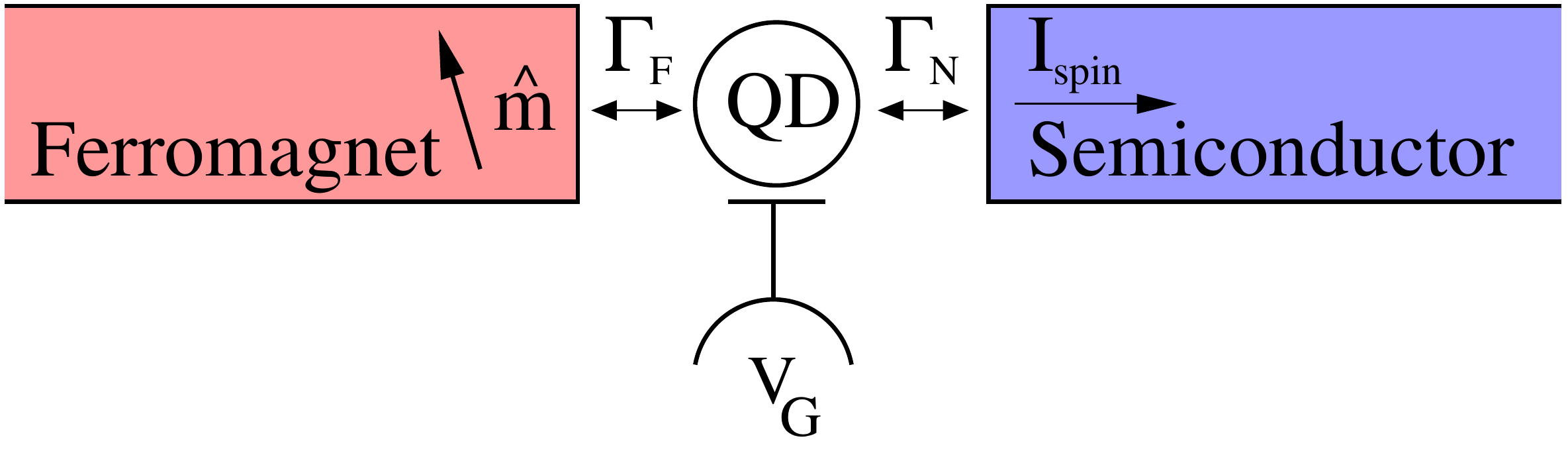}
\titlefigurecaption{%
  Setup of a device for generating a spin current into a semiconductor by adiabatic or non-adiabatic pumping of electrons through an interacting quantum dot.
  }

\maketitle   

\section{Introduction}

A substantial aspect of semiconductor spintronics~\cite{rev-zutic,rev-awschalom,rev-sinova} is the controlled generation
and manipulation of spin-polarized carriers in semiconductors. 
Both in the context of nanoscopic spintronic devices and quantum computation, it is highly desirable to coherently manipulate a few or even individual spins.
The main objective of this article is to review some recent studies of adiabatic charge and spin pumping from a spin-polarized lead (e.g. a ferromagnetic metal, a diluted magnetic semiconductor, or a spin-polarized edge channel of a quantum Hall system) through a quantum dot with large charging energy into a (nonmagnetic) semiconductor.
The ferromagnet and the semiconductor are considered as electron reservoirs.
For the ferromagnet, the density of states is spin dependent. 
In the case in which the average single-particle level spacing in the quantum dot exceeds all other energy scales relevant to transport, such as bias voltage and temperature, only one or a few orbital levels in the quantum dot participate in transport.
Coulomb interaction is accounted for by a charging energy, which typically dominates over all other energy scales in the system.
Transport can occur by tunneling between quantum dot and leads. 
For establishing pumping through the quantum dot, some of the system parameters, such as the strength of the tunnel couplings, tunable by split gates in electrostatically-defined quantum dots, the level position, tunable by
a gate voltage, an external magnetic field, or the magnitude and direction of the ferromagnet's magnetization, 
should vary in time, which makes the Hamiltonian explicitly time dependent.
For adiabatic, i.e. slow, pumping at least two varying parameters are necessary~\cite{brouwer}.

In the case of weakly interacting mesocsopic systems a theoretical description of pumping can rely on the scattering formalism~\cite{brouwer,buttiker-94,zhou-99,moskalets-01,makhlin-01,moskalets-02,entin-wohlman}. 
Quantum dots, however, are subject to strong Coulomb interaction. 
For interacting systems, only a few approaches~\cite{aleiner-98,aono-04,cota-94,janineprl,oreg,janineprb.06,silva,braun_burkard,citro,hernandez,yuge,deus} suited to specific systems or regimes exist. 
Therefore, we have developed~\cite{janineprb.06} a real-time diagrammatic technique to address charge and spin pumping through interacting quantum dots.
Pumping experiments have been realized both in open systems \cite{switkes99,giazotto} and in Coulomb-blockaded ones \cite{pothier92,watson_experimental_2003,blumenthal07,buitelaar08,maisi09,mirovsky10}.

\section{Model}

In general, the Hamiltonian of the systems that we study in this article can be written as a sum of three contributions as
\begin{equation}
\label{Hamiltonian}
  	H= H_{\text{dot}}+H_{\text{leads}}+H_{\text{tun}} \, ,
\end{equation}
where $H_{\text{dot}}$ describes the isolated dot, $H_{\text{leads}}$ the leads and  $H_{\text{tun}}$ the tunnel coupling between the dot and the leads.  
In this article we consider two models  for the dot [Eqs. (\ref{eq_H_dot})  and  (\ref{HDotMatrixGen})]:  either the dot has a single spin-degenerate orbital level (minimal model exhibiting non trivial interaction effects) or two spin-degenerate orbital levels (minimal model to include spin-orbit coupling).
Extensions to more complex systems such as double dots \cite{riwar2010} are straightforward.
\subsection{Single-level quantum dot}
When only one orbital level in the dot participates in transport, the dot can be described by the Anderson model as 
\begin{equation}
\label{eq_H_dot}
	H_{\text{dot}}=\sum\limits_{\sigma} \epsilon\, d^\dagger_{\sigma}\, d_{\sigma} + U\, n_\uparrow	\, n_\downarrow \,,
\end{equation}
where $\epsilon$ is the spin-degenerate orbital level, $U$ the on-site Coulomb repulsion, $d_\sigma$ $(d_\sigma^\dagger)$ the annihilation (creation) operator for an electron of spin $\sigma$, and $n_\sigma$ the corresponding number operator. 
We consider two leads, one normal and one ferromagnetic, modeled by reservoirs of non-interacting electrons with Hamiltonians 
\begin{align}
\label{eq_H_leads}
	H_{\text{lead\, N}}&=\sum\limits_{k,\sigma} \epsilon _{k}\,c^\dagger_{k\sigma}\,c_{k\sigma}\\ 
	H_{\rm{lead\, F}}&=\sum\limits_{k,\alpha} E_{k \alpha}a^\dagger_{k \alpha}a_{k \alpha} \,, 
\end{align}
where $c_{k\sigma }$  ($c^{\dagger}_{k\sigma }$) is the annihilation (creation) operator for an electron with spin $\sigma$ and wave vector $k$ in the normal lead, while $a_{k\alpha }$ ($a^{\dagger}_{k\alpha }$) is the annihilation (creation) operator of an electron with majority/minority spin $\alpha=\pm$ and wave vector $k$ in the ferromagnetic lead. 
In general, the majority and minority spin direction in the ferromagnet ($\alpha=\pm$) will be different from the spin quantization axis chosen for the dot and the normal lead ($\sigma=\uparrow,\downarrow$). 
The majority and minority creation operators are connected to the ones for electrons with spin $\sigma$ along the dot's quantization axis via the transformation $a_{k\alpha}^\dagger = \sum_\sigma A_{\alpha \sigma} a_{k\sigma}^\dagger$ with
\begin{equation}
\label{maj/min}
	\left( A_{\alpha \sigma} \right) = \left( 
	\begin{array}{cc}
	e^{- i \varphi/2 } \cos (\theta/2) & e^{i \varphi/2 } \sin(\theta/2)
	\\ 
	- e^{- i \varphi/2 } \sin(\theta/2) & e^{i \varphi/2 } \cos(\theta/2) 
	\end{array}
	\right)	
	\, ,
\end{equation} 
where the polar angle $\theta$ and azimuthal angle $\varphi$ define the ferromagnet's magnetization direction $\hat{e}_p= \left( \sin\theta \, \cos\varphi , \sin\theta \, \sin\varphi , \cos\theta \right)^\text{T}$ in the (time-independent) coordinate system with the $z$-axis chosen along the spin-quantization axis of quantum dot and normal lead. 
We define the ferromagnet polarization $p=(\rho_{\rm{F}+}-\rho_{\rm{F}-})/(\rho_{\rm{F}+}+
\rho_{\rm{F}-})$, where $\rho_{\rm{F}\alpha}$ are the densities of states at the Fermi energy for the majority and minority bands.
Finally, the tunnel coupling between dot and leads conserves spin and is described by 
\begin{equation}
\label{eq_H_tun}
	H_{\text{tun}}=\sum\limits_{k,\sigma}\left( V_{\rm{N}}\; c^\dagger_{k\sigma}\,d_{\sigma} +V_{\rm{F}}\, a^\dagger_{k\sigma}	\,d_{\sigma} +\text{h.c.}\right)\,. 
\end{equation}
The tunnel-matrix elements $V_{\rm{N}}$ and $V_{\rm{F}}$ 
define the tunnel-coupling strength to the normal and ferromagnetic lead by $\Gamma_{\rm N}=2 \pi \rho_{\rm{N}} \left|V_{\rm N}\right|^2$ 
and $\Gamma_{\rm {F}}=\sum_{\alpha} \pi \rho_{\rm{F}\alpha} \left|V_{\rm{F}}\right|^2$. 
The prefactor $1/2$ in the ferromagnetic case accounts for an average of majority and minority spin channels. 

We  use the single-level dot model described above with $\theta=0$ to study spin and charge through a quantum dot tunnel-coupled to a normal lead and a ferromagnetic lead with fixed (time-independent) magnetization direction in Secs.~\ref{sec:FDN} and \ref{sec:NA}.  
The full model with finite values of $\theta$ and $\varphi$ becomes relevant for a time-dependent magnetization direction, which we discuss in Sec.~\ref{sec:RM}. 

\subsection{Two-level quantum dot}
To investigate the effect of spin-orbit coupling, we need to consider at least two levels in the dot. The minimum dot's model that allows for time-reversal symmetry is 
\begin{align}
H_{\text{dot}}=	\Psi^\dagger\begin{pmatrix}(\epsilon_1-\frac{U}{2}) \bm{\sigma}_0 & -i \bm{\alpha}_{\rm so} \cdot \bm{\sigma} \\i \bm{\alpha}_{\rm so} \cdot \bm{\sigma} & (\epsilon_2-\frac{U}{2})\bm{\sigma} _0\end{pmatrix} \Psi+\frac{1}{2}
U(\Psi^\dagger\Psi)^2
\label{HDotMatrixGen} \, ,
\end{align} 
where $\Psi=(d_{1\uparrow},d_{1\downarrow},d_{2\uparrow},d_{2\downarrow})^{T}$, with $d_{\eta\sigma}$ being the annihilation operators for an electron with spin $\sigma$ in orbital level $\eta=1,2$. 
The entries of the $2 \times 2$ matrix, that reflects the Hilbert space of the two orbital levels, are operators in spin space: $\bm{\sigma}$ denotes the vector of Pauli matrices, $\bm{\sigma}_0$ is the identity matrix, and $\bm{\alpha}_{\rm so}$ is a real vector describing the spin-orbit coupling. 
The single particle energies $\epsilon_1$ and $\epsilon_2$ have been shifted by $U/2$ to guarantee that the energy difference between the empty-dot state and the dot's occupation by one electron is independent of the charging energy $U$. 
This allows us to study the limit of infinitely large $U$. Moreover, to simplify the discussion, we assume that the Coulomb repulsion between electrons on the same orbital level equals to the repulsion between electrons on level $\eta=1$ and $\eta=2$ (spin-independent).   

As spin-orbit coupling already breaks spin symmetry, we do not need to include a ferromagnet to obtain spin pumping.  In Sec.~\ref{sec:so}, we discuss the case of a two-level quantum dot coupled to two normal leads, each described by Eq.~\eqref{eq_H_leads}. 
For this case, the modified tunneling Hamiltonian~\eqref{eq_H_tun} reads:
\begin{equation}
	H_{\text{tun}}=\sum\limits_{k,\eta,\sigma}\left( V_{\rm{L}\eta}\; c^\dagger_{k\sigma\rm{L}}\,d_{\eta\sigma} +V_{\rm{R}\eta}\, c^\dagger_{k\sigma\rm{R}}	\,d_{\eta\sigma} +\text{h.c.}\right)\,.\label{eq:tunSO}
\end{equation}
The tunnel-coupling strength to the left/right (L/R) lead is given by an average with respect to the orbital degree of freedom, i.e., 
$\Gamma_\lambda=\sum_{\eta} \pi \rho_\lambda \left|V_{\lambda\eta}\right|^2$ for lead $\lambda= \text{L},\text{R}$. 

For both the two-level quantum dot and the single-level quantum dot with a ferromagnetic lead, the overall line width is the sum of the tunnel-coupling strengths to the lead, i.e., $\Gamma=\Gamma_{\rm{N}}+\Gamma_{\rm{F}}=\Gamma_{\rm{L}}+\Gamma_{\rm{R}}$.

We always assume a harmonic variation of the system parameters $\{X_1, X_2 \}$ in time, $X_1 (t) = \bar X_1 + \delta X_1 \sin (\Omega t )$ and $X_2 (t) = \bar X_2 + \delta X_2 \sin (\Omega t + \phi)$.
Pumping in the adiabatic regime occurs only for a finite phase shift $\phi$. 
In the limit of weak pumping, i.e., small amplitudes $\delta X_1$ and $\delta X_2$, the expressions for the pumped charge and spin, expanded up to bilinear order in $\delta X_1$ and $\delta X_2$, are proportional to $\sin \phi$
\cite{brouwer,buttiker-94,zhou-99}.

\section{Diagrammatic approach to adiabatic pumping}

The problem of theoretically addressing electronic transport through the quantum dot is complicated by the combination of (a) the large number of degrees of freedom in the leads, (b) the Coulomb interaction in the quantum dot, (c) the tunnel coupling between quantum dot and leads, and (d) the explicit time dependence of one or several parts of Eq.~(\ref{Hamiltonian}).
The strategy to deal with this problem is to perform a systematic perturbation expansion both in the tunnel-coupling strength, $\Gamma$, and the pumping frequency~\cite{janineprb.06}, $\Omega$, then to treat the leads as a reservoir of noninteracting degrees of freedom that can be integrated out with the help of Wick's theorem. 
In this way, one ends up with an effective description of the time evolution of a reduced system consisting of a few interacting degrees of freedom only. 

\subsection{Adiabatic expansion}

In order to perform the perturbation expansions we need first to identify the parts of the Hamiltonian that will be treated as perturbations.
For the expansion in the tunnel-coupling strength, this is the tunneling part of the Hamiltonian, while for the expansion in the pumping frequency, it is the time variation of the Hamiltonian that is considered to be small. 
This time variation of the Hamiltonian enters the evaluation of the quantum-statistical expectation value $\langle A(t)\rangle$ of any operator $A$ at time $t$ since the latter depends on the Hamiltonian $H(\tau)$ at earlier times $\tau$.
For a perturbative treatment, we expand $H(\tau)$ about the final time $t$ up to linear order,\footnote{How to include higher orders in the pumping frequency to study non-adiabatic pumping will be commented on in Section \ref{section-na}.} $H(\tau) \approx H(t) + (\tau-t) \dot{H}(t)$.
An additional complication can arise for a time-dependent quantum-dot Hamiltonian. 
Since we aim at formulating the diagrammatic language in the eigenbasis of the quantum-dot Hamiltonian, we may be forced to first employ a time-dependent unitary transformation $\mathcal{U}(t)$ such that $\bar H_{\text{dot}}:= \mathcal{U}^\dagger H_{\text{dot}} \mathcal{U}$ is diagonal.
Then, we split the transformed Hamiltonian $\bar H - i \mathcal{U}^\dagger \dot{\mathcal{U}}$ into the decoupled system $H_0$ frozen at the final time $t$ and the perturbation $V(\tau)$,
\begin{equation}
	H_t(\tau) \approx H_0 + V(\tau) \, ,
\end{equation}
where the index $t$ indicates the time about which the expansion is performed, and
\begin{eqnarray}
\label{H_0}
	H_0 &=& \bar H_{\text{dot}}(t) + H_{\text{leads}}(t)
\\
\label{V}
  	V(\tau) &=& \bar H_{\text{tun}}(t) + (\tau-t) \dot{\bar H}(t) -i \mathcal{U}^\dagger(t) \dot{\mathcal{U}}(t) 
\, .
\end{eqnarray}

Now, we are able to express the quantum-statistical expectation value of $A$ as an integral over the Keldysh contour $K$ from time $\tau=-\infty$ to $t$ and then back to $-\infty$,
\begin{equation}
\label{average-A}
	\langle A(t)\rangle = \text{tr} \left[ \varrho_0 T_K \exp \left( -i \int_K d\tau V(\tau)_I \right) A(t)_I \right] \, .
\end{equation}
The Keldysh time-ordering operator $T_K$ orders all operators along the Keldysh contour, the index $I$ indicates interaction picture with respect to $H_0$, and $\rho_0$ is the (full) density matrix of the decoupled system at time $-\infty$.
The latter is assumed to be a tensor product of the density matrices for the quantum dot and the leads, with the leads being in an equilibrium state with the magnetization direction being the same as at time $t$.
This procedure amounts to an approximation in the case of an explicit time dependence of the magnetization direction of the ferromagnet.
On the one hand, adiabatic corrections due to this time dependence in the lead are fully taken into account for the dynamics of the quantum dot's degrees of freedom.
On the other hand, nonequilibrium distributions in the ferromagnet, that arise due to the explicit time dependence already in the absence of the tunnel coupling to the quantum dot, are neglected.
This is justified as long as the time scale for relaxation towards equilibrium in the leads is much shorter than the time scale for transport, determined by tunneling.
In that case, the non-adiabatic corrections to the leads' density matrix are negligible in comparison to the non-adiabatic corrections of the transport processes.

\begin{figure*}
\begin{center}
\includegraphics*[width=4.5in]{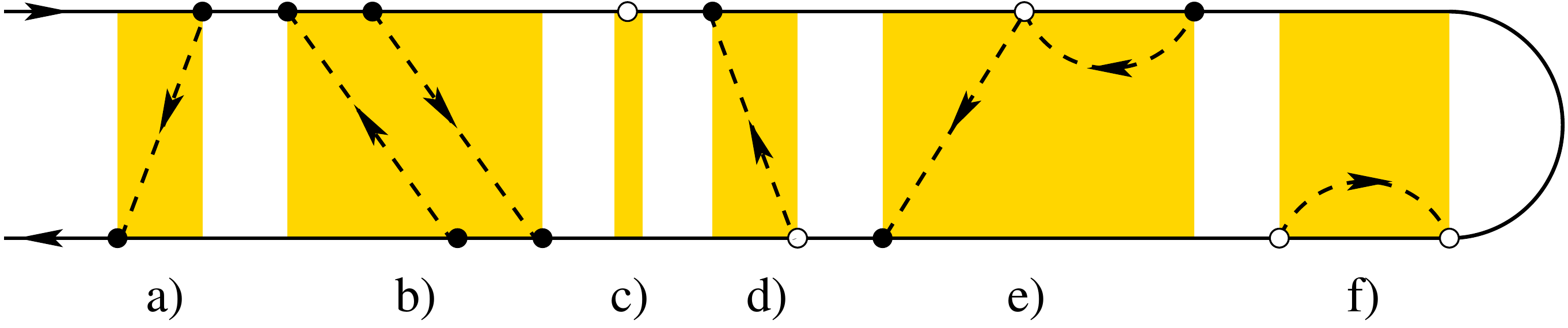}
\caption{Examples of irreducible blocks of diagrams. Filled vertices represent the tunneling Hamiltonian $\bar H_{\text{tun}}$ at time $t$. These are the only ones used for the instantaneous kernels shown for a) first order in $\Gamma$ and b) second order in $\Gamma$. Adiabatic corrections to the kernels contain one open vertex stemming from c) $\dot {\bar H}_{\text{dot}}$ or $-i \mathcal{U}^\dagger \dot{\mathcal{U}}$, d) $\dot {\bar H}_{\text{tun}}$ or e) $\dot H_{\text{leads}}$. Diagrams with f) two or more open vertices are higher order in the adiabatic expansion and, therefore, neglected for adiabatic pumping.
}  
\label{diagrams}
\end{center}
\end{figure*}

We proceed by expanding the exponential in powers of $V(\tau)_I$. 
The diagrammatic representation of each $V(\tau)_I$ is a vertex.
There are, in general, three different types of vertices, characterized by the number of lead operators being involved: this number may be zero for vertices originating from $\dot {\bar H}_{\text{dot}}$ or $-i \mathcal{U}^\dagger \dot{\mathcal{U}}$, one stemming from $\bar H_{\text{tun}}$ or $\dot {\bar H}_{\text{tun}}$, and two for vertices representing $\dot H_{\text{leads}}$.
In the next step, we integrate out the leads' degrees of freedom.
Using Wick's theorem, we contract the lead operators in pairs.
Each contraction is diagrammatically represented by a tunneling line.
As a consequence, the different types of vertices differ by the number of tunneling lines (zero, one or two) that are connected with the vertex.
The time evolution from $-\infty$ to $t$ is diagrammatically represented by an infinite sequence of irreducible blocks, defined as the parts of a diagram, where a vertical line at any time $\tau$ crosses at least one tunneling line. 
This infinite series of irreducible blocks can be summed up by a Dyson equation.
Examples of diagrams are depicted in Fig.~\ref{diagrams}.

The full density matrix $\varrho_\text{full}(t)$ of the coupled system includes both the quantum-dot and the leads' degrees of freedom.
By performing a partial trace over the latter, we arrive at the reduced density matrix $\varrho (t)= {\text{tr}}_{\text{leads}} [\varrho_\text{full}(t)]$, whose dynamics is described by the kinetic equation
\begin{equation}
\label{eq_rdm}
	\dot \varrho \left( t\right) + i [\bar H_{\text{dot}} -i \mathcal{U}^\dagger \dot{\mathcal{U}},  \varrho ]\left( t\right) = \int\limits_{-\infty}^{t}dt' \; W \left( t,t'\right)\varrho  \left( t'\right) \, .
\end{equation}
The commutator on the left hand side describes the coherent time evolution of the quantum dot in the absence of leads. The Liouville superoperator $W$ acting on the density matrix on the right hand side corresponds to the sum of all irreducible blocks.
The charge and spin currents $I_Q$ and $I_\mathbf{S}$ into a specified lead are obtained by
\begin{equation}
\label{eq_currents}
	I_{Q(\mathbf{S})} =
	\text{tr}_{\text{dot}} \left[ \int\limits_{-\infty}^{t}dt' \; W^{Q(\mathbf{S})} \left( t,t'\right)\varrho  \left( t'\right) 
	\right] \, ,
\end{equation}
where $W^{Q(\mathbf{S})}$ differs from $W$ by counting the charge/spin being transferred into the specified lead within the irreducible block.\footnote{The bold-face notation $\mathbf S$ indicates that spin and, thus, the spin current is, in general, a vector quantity.}

Although we have already linearized the time dependence of the Hamiltonian $H(\tau)$ about the time $t$, we still need to perform a systematic adiabatic expansion for the kinetic equation and the expression for the currents.
For this, we make use of the expansion $\varrho \left( t'\right) \approx \varrho \left( t\right) + (t'-t) \dot \varrho \left( t\right)$ on the right hand side of Eqs.~(\ref{eq_rdm}) and (\ref{eq_currents}).
Furthermore, we expand the kernel $W(t,t') \approx W^{(\text{i})}_t (t-t') + W^{(\text{a})}_t (t-t')$.
The instantaneous contribution, labeled by $(\text{i})$, corresponds to freezing all parameters at their values at time $t$.
The adiabatic correction, labeled by $(\text{a})$, is obtained by keeping exactly one vertex stemming from $\dot{\bar{H}}$ or $-i \mathcal{U}^\dagger \dot{\mathcal{U}}$.
Similarly, we perform an adiabatic expansion for the reduced density matrix $\varrho  \approx \varrho_t^{(\text{i})} + \varrho_t^{(\text{a})}$.
Since both the instantaneous part and the adiabatic correction to the kernel $W$ are functions of the time difference $(t-t')$, it is convenient to introduce the Laplace transform $W_t^{(\text{i/a})} (z) = \int_{-\infty}^{t} dt'\, \exp [-z(t-t')] W_t^{(\text{i/a})}(t-t')$ in order to define its zero-frequency value $W_t^{(\text{i/a})} =W_t^{(\text{i/a})}\left(z\right)|_{z=0^+}$ and its first derivative $\partial W_t^{(\text{i})} = \partial W_t^{(\text{i})}\left(z\right)/\partial z |_{z=0^+}$.

Now, we are in the position to complete the adiabatic expansion of Eq.~(\ref{eq_rdm}).
The lowest ({\em instantaneous}) order describes the equilibrium situation when all parameters are frozen to their values at time $t$,
\begin{align}
\label{rho_i}
	 i [\bar H_{\text{dot}}, \varrho_t^{(\text{i})}] = W_t^{(\text{i})} \varrho_t^{(\text{i})} \, .
\end{align}
Together with $\text{tr} \left[ \varrho_t^{(\text{i})} \right] =1$, we can solve for $\varrho_t^{(\text{i})}$, and plug this into the next-order ({\em adiabatic}) correction to the kinetic equation that contains all contributions linear in the pumping frequency,
\begin{align}
\label{rho_a}
	\dot \varrho_t^{(\text{i})} + i [\bar H_{\text{dot}}, \varrho_t^{(\text{a})}] + [ \mathcal{U}^\dagger \dot{\mathcal{U}},  \varrho_t^{(\text{i})}] =
	\nonumber \\
	 W_t^{(\text{a})} \varrho_t^{(\text{i})} + W_t^{(\text{i})} \varrho_t^{(\text{a})} + \partial W_t^{(\text{i})} \dot \varrho_t^{(\text{i})}\, ,
\end{align}
to solve (while making use of $\text{tr} \left[ \varrho_t^{(\text{a})} \right] =0$) for $\varrho_t^{(\text{a})}$.

In a similar way, we can perform an adiabatic expansion of the currents. 
The instantaneous parts vanish since they correspond to the equilibrium currents calculated with all parameters frozen at time $t$.
The adiabatic correction is
\begin{align}
	I_{Q(\mathbf{S}),t}^{(\text{a})} =
	\text{tr}_{\text{dot}} \left[ W_t^{Q(\mathbf{S}) (\text{a})} \varrho_t^{(\text{i})} 
	+  W_t^{Q(\mathbf{S}) (\text{i})} \varrho_t^{(\text{a})} \right] \, .
\end{align}
By integration over one pumping cycle with period $\mathcal{T} = 2 \pi / \Omega$, we obtain the pumped charge and spin as $Q=\int_0^{\mathcal{T}}I_{Q,t}^{(\text{a})} dt$ and $\mathbf{S}=\int_0^{\mathcal{T}}I_{\mathbf{S},t}^{(\text{a})} dt$, respectively. 
This scheme can be generalized to describe adiabatic pumping in the presence of a finite DC bias voltage~\cite{reckermann,calvo}.
In that case, the adiabatically pumped charge and spin adds to the instantaneous contribution. 

\subsection{Expansion in the tunnel coupling}

On top of the adiabatic expansion, we perform a systematic perturbation expansion of the kinetic equation in powers of the tunnel-coupling strength, $\Gamma$.
For this, we expand the kernel $W_t^{(\text{i/a})} = \sum_{n=1}^\infty W_t^{(\text{i/a},n)}$, where the index $n$ labels the power in the tunnel-coupling strength. 
The expansion starts to first order.
There is no zeroth order since in the absence of tunneling the kernel has to vanish.
The expansion of the instantaneous part of the reduced density matrix $\varrho_t^{(\text{i})}= \sum_{n=0}^\infty \varrho_t^{(\text{i},n)}$ starts to zeroth order, such that the normalization condition $\text{tr}_\text{dot} \left[ \varrho_t^{(\text{i})} \right] =1$ can be satisfied.

For a systematic expansion of the kinetic equations Eqs.~(\ref{rho_i}) and (\ref{rho_a}), we need to analyze the commutator $[\bar H_{\text{dot}}, \varrho_t^{(\text{i/a})}]$.
By construction, diagonal elements of the reduced density matrix drop out.
Off-diagonal matrix elements, describing coherent superpositions, are only nonzero if the energy difference of the corresponding states is of the order of or less than the tunnel-coupling strength. 
Therefore, we count this energy difference as one order in $\Gamma$.
As a result, to lowest order in $\Gamma$, Eqs.~(\ref{rho_i}) and (\ref{rho_a}) simplify to
\begin{align}
\label{rho_i small gamma}
	 i [\bar H_{\text{dot}}, \varrho_t^{(\text{i},0)}] = W_t^{(\text{i},1)} \varrho_t^{(\text{i,0})} \, .
\end{align}
and
\begin{align}
\label{rho_a small gamma}
	 \dot \varrho_t^{(\text{i},0)} + i [\bar H_{\text{dot}}, \varrho_t^{({\text{a}},-1)}] = W_t^{(\text{i},1)} \varrho_t^{(\text{a},-1)} \, .
\end{align}
In order to match the power of $\dot \varrho_t^{(\text{i},0)}$, the expansion of the adiabatic correction to the reduced density matrix, $\varrho_t^{(\text{a})}= \sum_{n=-1}^\infty \varrho_t^{(\text{a},n)}$ has to start to minus first order in $\Gamma$.\footnote{When going to the limit of weak tunnel coupling, $\Gamma \rightarrow 0$, one needs to keep in mind that the adiabaticity condition, $\Omega \ll \Gamma$, has to remain fulfilled. 
Therefore, $\varrho^{(\text{a},-1)}_{t} \propto \Omega/ \Gamma$ does not diverge.}
Furthermore, all the terms of Eq.~(\ref{rho_a}) involving $\varrho^{(\text{i})}_{t}$ do not appear to lowest order in $\Gamma$.
The expressions for the charge and spin current can be expanded in $\Gamma$ in a similar way. 
Since the instantaneous kernel starts to first and the adiabatic correction to the density matrix to minus first order, the expansion of the currents begin to zeroth order in $\Gamma$.

The diagrammatic rules of how to calculate the instantaneous kernel and its adiabatic correction depends on specific model system.
We do not derive or state them here.
Instead, we rather review the results obtained for spin pumping in various cases.
For technical details we refer the reader to earlier publications cited in the following.

\section{Adiabatic spin pumping through quantum dots}

In the following, we focus on generating spin currents into a normal lead by adiabatic pumping through a quantum dot.
To achieve a finite spin current, spin symmetry needs to be broken.
This could be done in various ways. 
Here, we restrict ourselves only to the cases in which a ferromagnetic lead is attached or spin-orbit coupling is used. 
Other cases such as applying an external magnetic field~\cite{mucciolo_2002,fransson_2010,fransson_2013} are not considered. 
Experimental realizations of quantum-dot systems with ferromagnetic leads include self-assembled InAs quantum dots \cite{hamaya}, metallic nanoparticles \cite{deshmukh_2002,mitani_2008,fert_2009,davidovic_2010}, semiconductor nanowires \cite{hofstetter_2010}, carbon nanotubes \cite{schoenenberger_2005,lindelof_2008}, and molecular devices \cite{pasupathy_2004}.
Spin-orbit coupling was studied in various quantum-dot systems \cite{Nowack_tune_SOI_exp,Fasth_SO_in_QD_exp,Nilsson_SO_in_QD_exp,Katsaros_SO_in_QD_exp,Takahashi_SO_in_QD_exp,Lai_SO_in_QD} and shown to be controllable \cite{Nowack_tune_SOI_exp,Lai_SO_in_QD}.
For all the model systems analyzed in this paper, spin symmetry is broken in a uniaxial way, i.e., for each moment in time, there is a rotational spin symmetry about a special axis determined by the magnetization direction of the ferromagnetic lead or the direction of the spin-orbit field. 
We do not investigate more complicated setups with non-collinear magnetic structure.
But even for uniaxial breaking of spin symmetry, the spin-pumping mechanisms can be separated into two classes.
In the first one, the spin-symmetry axis does not depend on time, while for the second one, pumping is achieved by varying the direction of the spin-symmetry axis.

\subsection{Pumping schemes with fixed spin-symmetry axis}

In the presence of a ferromagnetic lead with a fixed magnetization direction or for a quantum dot with a time-independent spin-orbit field, the fixed spin-symmetry axis provides the natural spin quantization axis.
It is obvious that any pumped spin current possesses only a component along this axis. 
In the adiabatic regime, two time-dependent parameters are required to establish pumping.
In the following, we combine the variation of gate voltages determining the energies of the quantum-dot levels, a time dependence of the tunnel-coupling strength, and a variation of the magnetization amplitude of a ferromagnetic lead.
The time variation of gate voltages implies that, in general, both charge and spin are being pumped. 
Pure spin pumping, i.e., spin without charge pumping, can only be achieved (if at all) for special choices of the system parameters at which charge pumping happens to vanish.

\subsubsection{F-QD-N with fixed spin-symmetry axis\label{sec:FDN}}

In our first example, we consider a single-level quantum dot with level position $\epsilon$ and charging energy $U$ associated to double occupancy of the quantum dot, which is coupled to one ferromagnetic, $\Gamma_\mathrm{F}$, and one normal lead, $\Gamma_\mathrm{N}$. 
The Hamiltonian for this system is given by Eq. ~(\ref{Hamiltonian}), together with Eqs. (\ref{eq_H_dot})--(\ref{eq_H_tun}) with the polar angle of the magnetization $\theta$ set to zero.
We concentrate on the limit of weak tunnel coupling, for which only kernels to first order in $\Gamma$ need to be considered.
An explicit calculation~\cite{splettstoesser_2008,winkler_2013} shows that for the single-level quantum dot the lowest- (zeroth-) order contribution to the pumped charge and spin current into the normal lead can be expressed in terms of the average quantum-dot occupation $\langle n\rangle$ via
\begin{align}
\label{eq_charge_current}
	I_Q^{(\text{a},0)}(t) =& -e
	\frac{\Gamma \Gamma_{\mathrm{N}}}{\Gamma^2-p^2\Gamma_\mathrm{F}^2}
	\frac{d\langle n\rangle^{(\text{i},0)}}{d t}
	\\
\label{eq_spin_current}
	I_S^{(\text{a},0)}(t) =& \frac{\hbar}{2}
	\frac{p\Gamma_\mathrm{F}\Gamma_{\mathrm{N}}}{\Gamma^2-p^2\Gamma_\mathrm{F}^2}
	\frac{d\langle n\rangle^{(\text{i},0)}}{d t}
\ ,
\end{align}
where $e<0$ is the electron charge. 
Here, $I_S$ is the current of the spin projected on the symmetry axis and, therefore, a scalar quantity.
To zeroth order in the tunnel coupling, the instantaneous occupation probabilities of the quantum-dot states are determined by the Boltzmann factors of the associated energies.
This yields for the average dot occupation
\begin{equation}
	\langle n\rangle^{(\text{i},0)} = \frac{2f(\epsilon)}{1+f(\epsilon)-f(\epsilon+U)} \, .
\end{equation}
As a consequence, pumping can (to this order in the tunnel-coupling strength) only appear if the dot energy $\epsilon$ is varied in time via a gate voltage.
This fixes already one of the two pumping parameters.
The other one can be either the tunnel couplings to the normal or the ferromagnetic lead, or it can be the degree of spin polarization $p$ in the ferromagnet.

In any case, we immediately derive from Eqs.~(\ref{eq_charge_current}) and (\ref{eq_spin_current}) that at any time, the ratio between charge and spin current is given by
\begin{equation}
\label{sqfact}
\frac{I_S^{(\text{a},0)}(t)}{I_Q^{(\text{a},0)}(t)}
= -\frac{\hbar}{2e} \frac{p\Gamma_\mathrm{F}}{\Gamma} \, .
\end{equation}
This expression is always positive, implying that, at any time, spin and charge flow in the same direction, compatible with the notion that electrons carry charge and spin at the same time.

The situation is different when integrating over a full pumping cycle.
There are times during the cycle when the instantaneous charge current is positive and others when it is negative.  
These contributions partially (sometimes fully) compensate each other.
The uncompensated part yields the net pumped charge, which may be positive or negative (or zero).
While at any moment in time, the relative sign of the spin and charge currents is fixed, their relative magnitude, expressed by the right hand side of Eq.~(\ref{sqfact}), may be time-dependent. 
Therefore, the compensation between positive and negative contributions for the pumped spin is different than the one for the pumped charge.
As a result, the ratio of spin and charge \textit{pumped per period} can be both positive or negative.
A divergence of this ratio indicates pure spin pumping. 
In order to explicitly calculate this ratio in bilinear response, i.e. for an infinitesimal area in parameter space, we have to specify the choice of pumping parameters.
In the following, we define the spin efficiency as the ratio
\begin{equation}
	R = - \frac{2e}{\hbar} \frac{S}{Q} \, ,
\end{equation}
with $S=|\bm{S}|$.
If we choose $\left\{\epsilon,\Gamma_\mathrm{N}\right\}$ or $\left\{\epsilon,\Gamma_\mathrm{F}\right\}$ as pumping parameters, we obtain the spin efficiency \cite{splettstoesser_2008}
\begin{equation}
R_{\epsilon,\mathrm{N/F}}
\label{eq_charge_ratio}
 = - p \frac{ 1+p^2\left( \bar{\Gamma}_\mathrm{F} / \bar{\Gamma}\right)^2
-2\left( \bar{\Gamma}_\mathrm{F} / \bar{\Gamma}\right)
}{
1+p^2\left( \bar{\Gamma}_\mathrm{F} / \bar{\Gamma}\right)^2-
2p^2 \left( \bar{\Gamma}_\mathrm{F} / \bar{\Gamma} \right)
}\ .
\end{equation}
\begin{figure}
\begin{center}
\includegraphics[width=3.in]{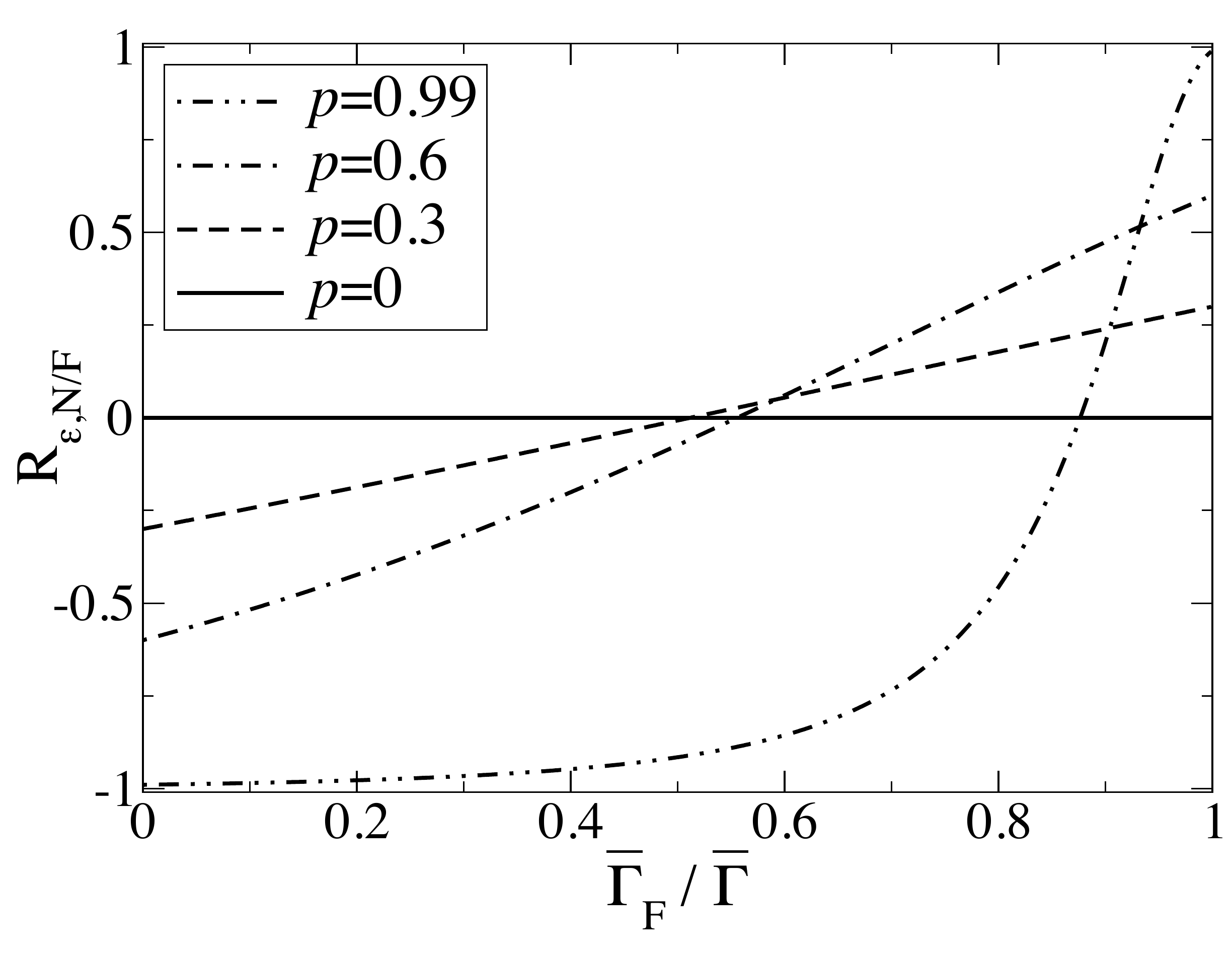}
\caption{
Spin efficiency $R_{\epsilon,\mathrm{N/F}}$ as a function of the relative tunnel-coupling strength to the ferromagnetic lead for different values of the polarization.  Figure adapted from Fig. 2(a) of Ref.~\cite{splettstoesser_2008}.
}  
\label{fig_ratio_p}
\end{center}
\end{figure}
It is shown in Fig.~\ref{fig_ratio_p} as a function of the relative coupling to the ferromagnetic lead. 
Obviously, the absolute value of the spin efficiency is one for the case of fully polarized lead, $p=1$ (not shown). In the more realistic case of $p<1$ the ratio 
$R_{\epsilon,\mathrm{N/F}}$ goes from $-p$ for vanishing coupling to the ferromagnetic lead ($\Gamma_{\text{F}}\rightarrow 0$) to $p$ for vanishing coupling to the normal lead ($\Gamma_{\text{N}}\rightarrow 0$). 
The sign change of $R_{\epsilon,\mathrm{N/F}}$ is an intriguing result, which implies that the relative direction of the pumped spin and charge can be controlled by the relative coupling to the normal and ferromagnetic lead. 

When pumping with the polarization of the ferromagnetic lead and the level position, $\left\{\epsilon,p \right\}$, we obtain
\begin{equation}
R_{\epsilon,p}
 = \frac{ 1+\bar{p}^2\left( \Gamma_\mathrm{F} / \Gamma \right)^2}
 {\bar{p}\left( \Gamma_\mathrm{F} / \Gamma\right)} \ .
\end{equation}
In this case, the spin efficiency is always positive. 

\paragraph{Pure spin currents}

Pure spin currents are realized when the pumped charge per cycle vanishes and, therefore, the spin efficiency $R$ diverges.
Such a divergence is not seen for any of the three choices $\left\{\epsilon,\Gamma_\mathrm{N}\right\}$, $\left\{\epsilon,\Gamma_\mathrm{F}\right\}$ or $\left\{\epsilon,p\right\}$ individually.
It can, however, occur when combining two of these pumping schemes, which is experimentally relevant whenever the variation of one external parameter affects two of the system parameters $\Gamma_\mathrm{N}$, $
\Gamma_\mathrm{F}$, and $p$. We now consider a time variation of the magnetization amplitude of the ferromagnet, $M(t)$.  
This can be realized by applying a small time-dependent magnetic field in a diluted magnetic semiconductor.  Microscopically this leads to a time-dependent splitting of the minority and majority bands. 
The majority and minority bands  $E_{k\pm}(t)$ are shifted in such a way to keep both the total number of electrons and the Fermi energy of the ferromagnet constant. This implies that  not only the polarization $p(t)$ varies in time  but also the tunnel-coupling  strength $\Gamma_\mathrm{F}(t)$. The time variations of $p(t)$ and $\Gamma_\mathrm{F}(t)$ are in phase and do not generate any pumping in the adiabatic limit. However, if the level position is also time dependent, the two pumping cycles $\left\{\epsilon,p\right\}$ and $\left\{\epsilon,\Gamma_\mathrm{F}\right\}$ occur simultaneously.
Their relative contributions to the total pumped charge and spin in the weak-pumping regime depend on the ratio 
\begin{equation}
	\nu=  \frac{ \delta \Gamma_{\mathrm{F}} / \bar {\Gamma}_{\mathrm{F}} } {\delta p / \bar p} \, ,
\end{equation}
which can be related to details of the ferromagnet's band structure (see Ref.~\cite{winkler_2013} for more details). Here it suffices to notice that $-1/2 \le \nu \le 0$ for parabolic bands, with $\nu\propto p^2$ for small polarizations. 
For the special value $\nu=0$ only the cycle $\left\{\epsilon,p\right\}$ is active. 
Positive values of $\nu$ are attainable for non-parabolic bands, for example when the density of states near the Fermi energy in the ferromagnet depends on energy as $\rho_{\text{F}}=(\rho_{\text{F}+}+\rho_{\text{F}-})/2 \propto\omega^{-d}$ with $d>0$. 

\begin{figure}
	\includegraphics[width=0.475\textwidth]{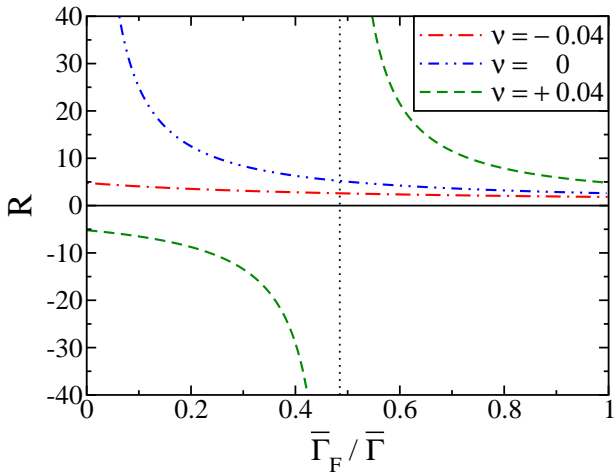} 
	\caption{(Color online) Total spin efficiency as a function of the relative tunnel-coupling strength to the ferromagnetic lead for $p=0.2$ and different values of $\nu$.
		   Figure taken from Fig. 7 of Ref.~\cite{winkler_2013}. }
	\label{fig_sum_ratio}
\end{figure}

We define the total spin efficiency for the two cycles as 
\begin{equation}
 	R=-\frac{2e}{\hbar}\frac{S_{\epsilon,p} + S_{\epsilon,\Gamma_\mathrm{F}}}{Q_{\epsilon,p} + Q_{\epsilon,\Gamma_\mathrm{F}}}  \, . 
\end{equation}
It is shown in Fig.~\ref{fig_sum_ratio} as a function of $\bar{\Gamma}_{\rm{F}}/\bar{\Gamma}$ for polarization $p=0.2$ and different values of $\nu$.
For $\nu=0$ (realized for flat bands around the Fermi energy), the spin efficiency diverges for $\bar{\Gamma}_{\rm{F}}/\bar{\Gamma} \rightarrow 0$, indicating a pure spin current, which is, however, only asymptotically reached as for $\bar{\Gamma}_{\rm{F}}/\bar{\Gamma} \rightarrow 0$ the spin current vanishes. For negative values of $\nu$ the divergence of $R$ is removed and a pure spin current is not realized. The most interesting case is $\nu>0$, when the condition for a pure spin current (divergence of $R$) is realized for a finite value of  $\bar{\Gamma}_{\rm{F}}/\bar{\Gamma}$ and hence when the spin current is of finite amplitude.

\subsubsection{N-QD-N with spin-orbit coupling\label{sec:so}}

Breaking of spin symmetry can also be achieved by spin-orbit coupling and, therefore, can lead to finite spin pumping~\cite{Sharma_SOI_spinpumping,Governale_SOI_spinpumping}. 
The minimal model that allows for a coupling of spin and orbital degrees of freedom is a two-level quantum dot, described by the Hamiltonian given by Eq.~\eqref{HDotMatrixGen}.
The matrix in Eq.~\eqref{HDotMatrixGen} has the most general form for a quantum dot with two orbital levels that allows for time-reversal symmetry and has been used recently to study transport phenomena in the presence of spin-orbit coupling \cite{brosco_prediction_2010,Eto_spin_filter,grap_interplay_2012,droste_josephson_2012}.

Here, we assume that the orbital levels, $\epsilon_{1}(t)$ and $\epsilon_{2}(t)$, can be changed in time.
Therefore, we consider in the following weak adiabatic pumping with $\{ \epsilon_1 , \epsilon_2 \}$, or, equivalently, with $\{ \epsilon , \Delta \epsilon \}$, where $\epsilon= (\epsilon_1 + \epsilon_2)/2$ is the mean level position and $\Delta \epsilon= (\epsilon_1 - \epsilon_2)/2$ half the level spacing.

To achieve pumping, the symmetry of the coupling to the left and the right lead needs to be broken.
Four real tunnel-matrix elements, $V_{\lambda \eta}$, of the tunneling Hamiltonian given by Eq.~\eqref{eq:tunSO} determine the coupling configuration of the two-level quantum dot to the leads. For further discussion, it is convenient to distinguish the ratio of the tunnel-coupling strength to the left lead and to the right lead, $\Gamma_\lambda$, from the difference between the coupling of the two orbital levels to the same lead. 
The latter is characterized by angles $\vartheta_\lambda$, which parametrize the tunnel-matrix elements via the relations $V_{\lambda 1}=\sqrt{\frac{\Gamma_{\lambda}}{\pi\rho}} \cos\frac{\vartheta_\lambda}{2}$ and $V_{\lambda 2}=\sqrt{\frac{\Gamma_{\lambda}}{\pi\rho}} \sin\frac{\vartheta_\lambda}{2}$. 
The pumped charge and spin are, in principle, finite if the left-right symmetry is broken by $\vartheta_\text{L} \neq \vartheta_\text{R}$. 
To break the symmetry via $\Gamma_\text{L}\neq\Gamma_\text{R}$ while keeping $\vartheta_\text{L} =\vartheta_\text{R}$ is not sufficient to establish finite pumping \cite{rojek} since this case corresponds to an effective one-parameter pumping with vanishing pumped charge and spin in the adiabatic regime. 
For simplicity, we restrict ourselves in the following to the case $\Gamma_\text{L}=\Gamma_\text{R}$.

To comment on the influence of Coulomb interaction, we compare the limit $U\rightarrow\infty$, to the limit of vanishing charging energy, $U=0$. In both regimes, the absolute value of the pumped spin 
is maximal
for small but finite spin-orbit coupling, at approximately $\alpha_{\rm so}= \Gamma/10$~\cite{rojek}. The results for the spin efficiency, discussed in the following, are calculated for this value in the weak coupling regime, i.e., lowest (zeroth) order $\Gamma$.

\begin{figure}
	\includegraphics[width=.45\textwidth]{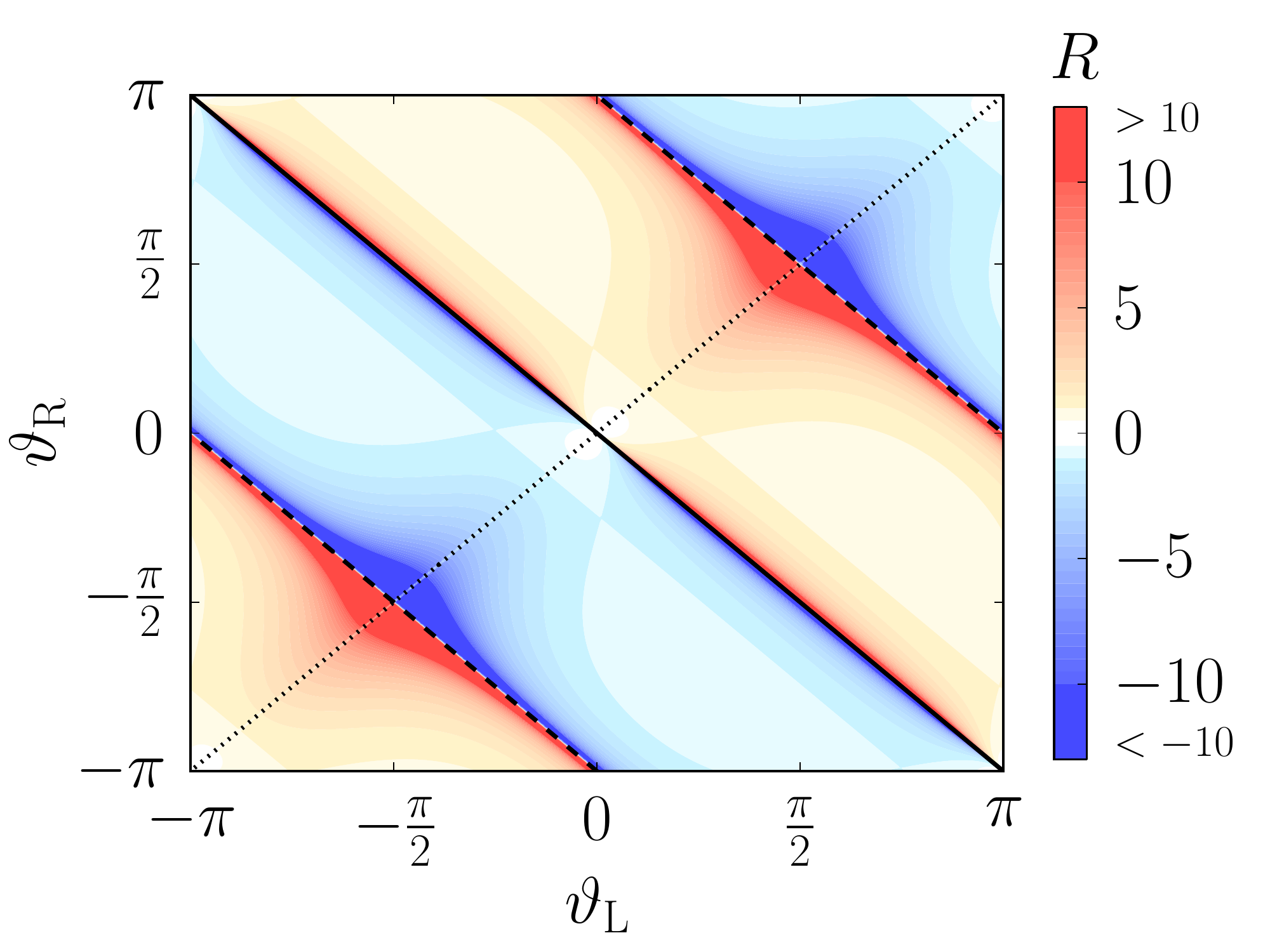}
	\caption{\label{fig:SOU0}(Color online) Spin efficiency in the $U=0$ limit as a function of $\vartheta_L$ and $\vartheta_R$ for fixed $\Gamma_{\text{L}}=\Gamma_{\text{R}}$ and $\alpha_\text{so}=\Gamma/10$. 
	The dotted line indicates left-right symmetric tunnel couplings with vanishing pumped charge and spin. 
	Along the solid ($V_{\text{L}1}/V_{\text{L}2}=-V_{\text{R}1}/V_{\text{R}2}$) and dashed ($V_{\text{L}1}/V_{\text{L}2}=V_{\text{R}2}/V_{\text{R}1}$) lines the spin efficiency diverges, indicating a pure spin current. 
	}
\end{figure}

\begin{figure}
	\includegraphics[width=.45\textwidth]{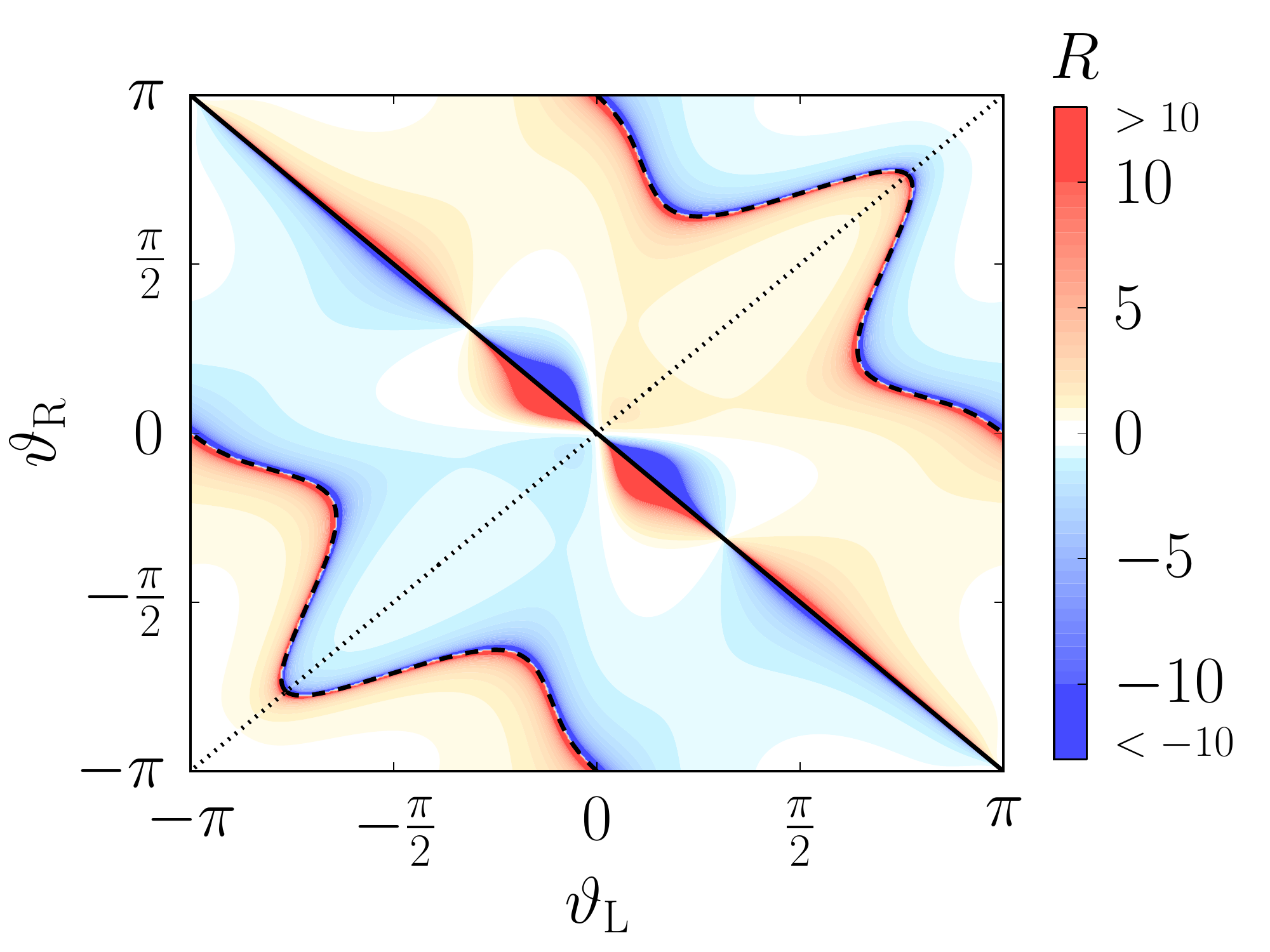}
	\caption{\label{fig:SOUI}(Color online) Spin efficiency in the $U=\infty$ limit as a function of $\vartheta_L$ and $\vartheta_R$ for fixed $\Gamma_{\text{L}}=\Gamma_{\text{R}}$ and $\alpha_\text{so}=\Gamma/10$. 
	The dotted line indicates left-right symmetric tunnel couplings with vanishing pumped charge and spin. 
	Along the solid ($V_{\text{L}1}/V_{\text{L}2}=-V_{\text{R}1}/V_{\text{R}2}$) and dashed lines the spin efficiency diverges, indicating a pure spin current.
	The position of the dashed lines depends on the values of $\overline\epsilon$ and $\overline{\Delta \epsilon}$.
	Here, we choose $\overline{\epsilon}=k_BT$ and $\overline{\Delta\epsilon}=2 \Gamma/10$. 
	}
\end{figure}

Figures~\ref{fig:SOU0} and \ref{fig:SOUI} 
show spin efficiency in the $U=0$ and $U\rightarrow\infty$ limit, respectively, depending on the coupling configuration parametrized by $\vartheta_\text{L}$ and $\vartheta_\text{R}$. 
The dotted lines indicate the case of left-right symmetric tunnel couplings, $V_{\text{L}\alpha}=V_{\text{R}\alpha}$, where both the pumped charge and spin vanish. The possibility of pure spin pumping in case of other coupling configurations shall be the focus of the remaining lines of this section.

We start with the discussion of the spin efficiency in the $U=0$ limit, depicted in Fig.~\ref{fig:SOU0}. 
One important feature is the fact that the spin efficiency is independent of the time-averaged level positions, determined by $\overline{\epsilon}$ and $\overline{\Delta \epsilon}$. 
Thus, the pumped charge and spin behave similarly under exchange of the orbital levels.
Particle-hole symmetry implies that both pumped charge and spin are antisymmetric under $(\overline{ \epsilon}, \overline{\Delta \epsilon}) \rightarrow (-\overline{ \epsilon}, -\overline{\Delta \epsilon})$.
To lowest order in $\Gamma$, it turns out that the pumped charge and spin are symmetric under $\overline \epsilon\rightarrow -\overline \epsilon$, i.e., antisymmetric under $\overline {\Delta \epsilon} \rightarrow -\overline {\Delta \epsilon}$ alone.
As a consequence, pumped charge and spin both vanish for $\overline {\Delta \epsilon}=0$.
Apart from that situation of degenerate orbital levels and special coupling configurations, pumped charge and spin are finite and their ratio does not depend on the levels' positions.
On the other hand,  the spin efficiency does depend on the coupling configuration, $\vartheta_\text{L}$ and $\vartheta_\text{R}$, as illustrated by Fig.~\ref{fig:SOU0}. There are even two coupling configurations where spin efficiency diverges indicating a pure spin current. First, the dashed lines in Fig.~\ref{fig:SOU0} which show a divergence along and a large spin efficiency next to coupling configurations given by the relation  $V_{\text{L}1}V_{\text{R}1}=V_{\text{L}2}V_{\text{R}2}$. However, comparison with calculations for $U=0$ which are exact in $\Gamma$, e.g., by means of a scattering matrix approach \cite{brouwer,buttiker-94,brosco_prediction_2010,avron_geometry_2000}, show that the pumped charge vanishes only in lowest order in $\Gamma$. It also becomes finite for 
$U\rightarrow\infty$ (and lowest order in $\Gamma$) and, thus, this parameter configuration is not a 
good candidate to achieve  pure spin pumping. Along the solid line, on the other hand, where one level is symmetrically and the other one antisymmetrically coupled to the left and right lead, i.e. $V_{\text{L}1}=V_{\text{R}1}$ and $V_{\text{L}2}=-V_{\text{R}2}$ (or equivalently $1\leftrightarrow 2$), the coupling configuration leads to a pure spin current not only for calculations in lowest but to all orders in $\Gamma$ \cite{brosco_prediction_2010}. This pure spin current is independent of $\overline\epsilon$ and $\overline{\Delta \epsilon}$. 

Until now we have neglected the Coulomb interaction on the quantum dot although it is often large in few-electron devices. 
The question arises whether and how the condition for pure spin currents are modified due to Coulomb interaction.
The following discussion focuses on the limit of an infinitely large charging energy. 

In the $U\rightarrow\infty$ limit the antisymmetry with respect to $\overline {\Delta \epsilon}$ is broken. 
This can be understood by introducing an isospin associated with the orbital degree of freedom (where isospin up and down correspond to an occupation of orbital level 1 and 2, respectively). 
The Coulomb repulsion introduces an effective exchange field which acts on this isospin \cite{rojek} and leads to the breaking of the aforementioned antisymmetric behavior.
To calculate this exchange field, a high-energy cutoff parameter, $U_\text{cutoff}$, for the energy integration is used to guarantee convergence. 
It affects the amplitude of the exchange field and is physically given by the smaller of the band width of the leads and the charging energy \cite{schoeller_mesoscopic_1994,knig_resonant_1996,knig_zero-bias_1996} (we set $U_\text{cutoff}=100 k_BT$). By the influence of the isospin exchange field, the condition of vanishing pumped charge and spin are shifted away from $\overline {\Delta \epsilon}=0$. More importantly, this shift is different for the pumped charge and the pumped spin. As a consequence, pure spin pumping can be achieved by fine tuning of the orbital levels. 
Viewed in an alternative way, there are, for fixed orbital levels, coupling configurations that support a pure spin current.
They depend significantly on the orbital levels. 
As an example, we chose $\overline{\epsilon}=k_BT$ and $\overline{\Delta\epsilon}=2 \Gamma/10$, which leads to pure spin currents along the dashed lines in Fig.~\ref{fig:SOUI}.
Note that the energy scale on which spin efficiency changes is different for $\overline{\epsilon}$ and $\overline{\Delta \epsilon}$. For the time-averaged mean level position, $\overline{\epsilon}$, the temperature provides the energy scale. On the other hand, the time-averaged level spacing, $\overline{\Delta \epsilon}$ (as well as $\alpha_\text{so}$) has to be compared to the tunnel-coupling strength, $\Gamma$.     

As a much stronger statement, there are cases in which pure spin current is not only possible for special, fine-tuned orbital energies but for {\em all} values of $\overline \epsilon$ and $\overline {\Delta \epsilon}$. 
The coupling configuration $\vartheta_\text{L}=-\vartheta_\text{R}$, i.e. $V_{\text{L}1}=V_{\text{R}1}$ and $V_{\text{L}2}=-V_{\text{R}2}$ (or equivalently $1\leftrightarrow 2$), represented by the solid line, leads universally to a pure spin current \cite{rojek}. This generalizes the result found for $U=0$ \cite{brosco_prediction_2010} to the limit of strong Coulomb interaction. 
While we have focused on the weak-coupling regime, it has been shown recently that at this coupling configuration pure spin currents remain even in the strong-coupling limit reached at very low temperatures \cite{moca_2013}.

\subsection{Pumping with rotating lead magnetization\label{sec:RM}}

So far, we have considered spin pumping schemes in which either the magnetization direction of a ferromagnetic lead or a spin-orbit field defined a rotational spin-symmetry axis that was not changed in time.
We now turn to the case in which the direction of the spin-symmetry axis changes in time.
To be more specific, we consider a F-QD-N setup in which the magnitude $M$ of the ferromagnet's magnetization $\vec{M}(t)=M {\hat{e}}_p(t)$ remains fixed but its direction ${\hat{e}}_p(t)= \left( \sin\theta \, \cos\varphi(t), \sin\theta \, \sin\varphi(t), \cos\theta \right)^\text{T}$ rotates about the $z$-axis. 
The fixed rotation axis $z$ is used as the (time-independent) quantization axis for the electron spins $\uparrow$ and $\downarrow$ in the dot and normal lead. On the other hand, the direction of the majority and minority spins $+$ and $-$ in the ferromagnet  varies in time. The Stoner splitting $\Delta E = E_{k-} - E_{k+}$ remains constant in time.
Such a rotation of the ferromagnet magnetization can be induced experimentally by means of  ferromagnetic resonance. 
The described mechanism is similar to the idea of recently proposed spin batteries without involving quantum dots~\cite{tserkovnyak_enhanced_2002,brataas_spin_2002,tserkovnyak_spin_2002,mahfouzi_microwave_2010,costache_electrical_2006}.
It is, furthermore, related to the proposal of charge pumping induced by the rotation of the magnetization direction of a magnetic quantum dot that is placed in between a ferromagnet and a normal lead~\cite{bender_tserkovnyak_brataas_2010}.

The two independent parameters necessary for adiabatic pumping to take place are  the  $x$- and $y$-components of the polarization, $p_x(t)=p \sin \theta \cos \varphi (t)$ and $p_y(t)=p \sin \theta \sin \varphi (t)$. 
In contrast to the pumping schemes discussed so far, no time-dependence of the level position is required to achieve pumping.  

The fact that only magnetic properties associated with one of the leads are varied in time has important consequences.
First, the lowest- (zeroth-) order contribution to charge and spin pumping vanishes. 
This follows from Eqs.~(\ref{eq_charge_current}) and (\ref{eq_spin_current}) since the instantaneous average dot occupation $\langle n \rangle^{(i,0)}$ is constant in time.
It is, therefore, necessary to include the next-order contribution in the perturbation expansion in the tunnel-coupling strength.
The second consequence is that, even to higher order in the tunnel-coupling strength, the pumped charge remains zero because spin and charge are decoupled. 
But this, in turn, means that there is always pure spin pumping without the complication to tune gate voltages or tunnel-coupling strengths.

An explicit calculation \cite{winkler_2013} yields that the pumped spin current can be nicely written in a compact analytical form,
\begin{align}
I_{\mathbf{S},t}^{(\text{a},1)} &= \frac{G_0}{8\pi} \left\{ A  \left( {\hat{e}}_p  \times \partial_t {\hat{e}}_p  \right) +(1-A)  \partial_t {\hat{e}}_p  \right\}   \,. 
\label{scur_a_1}
\end{align}
Here, we made use of the dimensionless linear conductance (in units of $e^2/h$)
\begin{equation}
	G_0= -2\pi \;\frac{\Gamma_\text{N} \; \Gamma_\text{F}}{\Gamma^2} \;  \frac{\partial_\epsilon    \langle n \rangle^{(\text{i},0)}}{\tau^Q_\text{rel}}  \;   
\end{equation}
of the F-dot-N structure for vanishing polarization.  We have also defined the enhancement factor
\begin{equation}
\label{enhancement}
	A =
	1+\frac{\Gamma_\text{N}}{\Gamma_\text{F}} \frac{\left(B \tau^S_\text{rel}\right)^2}{1 +  \left(B \tau^S_\text{rel}\right)^2 } \, ,     
\end{equation}
where $\tau^Q_\text{rel}$ and $\tau^S_\text{rel}$ are the charge and spin relaxation times, respectively, and $B$ describes an interaction-induced exchange field that is a consequence of the spin-dependent tunnel coupling of the dot level to the ferromagnet~\cite{koenig_interaction_2003,martinek_kondo_qd_2003,braun_theory_2004}. 
The exchange field always appears in combination with the spin relaxation time $\tau^S_\text{rel}$, as this is the time scale during  which the field can act on the quantum-dot spin before it relaxes due to tunneling events with the leads.  

The explicit expressions for the relaxation rates and the exchange field read
\begin{align}
\frac{1}{\tau^Q_\text{rel}} 
&= \Gamma \left[1 + f(\epsilon) - f(\epsilon+U)  \right] 
\\
\frac{1}{\tau^S_\text{rel}} 
&= \Gamma \left[1 - f(\epsilon)+ f(\epsilon+U)  \right] 
\\
B &=  \frac{\Gamma_{\text{F}} \, p }{\pi} \; \mathcal{P}\!\!\!\!\!\!\int\!\! d\omega  \left[ \frac{1 - f(\omega)}{ \omega - \epsilon }  \;+\;   \frac{f (\omega) }{ \omega - \epsilon - U } \right] \, ,
\end{align}
where 
$\mathcal{P}\!\!\!\!\!\!\int\!\! \,\, d\omega$ denotes Cauchy's principal value.

The pumped spin is obtained by  integrating the spin current Eq.~(\ref{scur_a_1}) over one cycle.  Due to the symmetry of the problem, the pumped spin per cycle has only a component along the $z$-axis. 
Since $\partial_t {\hat{e}}_p$ does not have any $z$-component, it is only the term proportional to ${\hat{e}}_p \times \partial_t {\hat{e}}_p$ that contributes to the finite pumped spin per cycle $S_\varphi$.
Making use of  $({\hat{e}}_p \times \partial_t {\hat{e}}_p) \cdot \hat e_z = \dot\varphi(t) \sin^2 \theta $ and assuming a constant angular velocity, $\Omega= \dot{\varphi}(t)$,
we obtain for the pumped spin
\begin{align}
S_\varphi = \frac{1}{4} \Omega \sin^2\theta \,G_0 A \, .  
\label{eq_spin_current_a1}
\end{align}
The pumped spin is proportional to the enhancement factor $A$. It is interesting to notice that the presence of the exchange field leads to  $A>1$ and thus to a larger pumped spin. 
We will give here a qualitative picture of how this occurs. The exchange field $B$ affects the spin dynamics in two ways. It induces a spin precession of the spin accumulated on the dot along $\hat e_p$ thus reducing the pumped spin. Additionally, it affects the spin-accumulation processes. This latter effect dominates and gives rise to an overall enhancement of the pumped spin. 
Due to the prefactor $\Gamma_\text{N}/\Gamma_\text{F}$ in the second term of Eq.~(\ref{enhancement}),  the exchange field becomes more important when increasing the tunnel coupling to the normal lead.
 
\begin{figure}
	\includegraphics[width=0.45\textwidth]{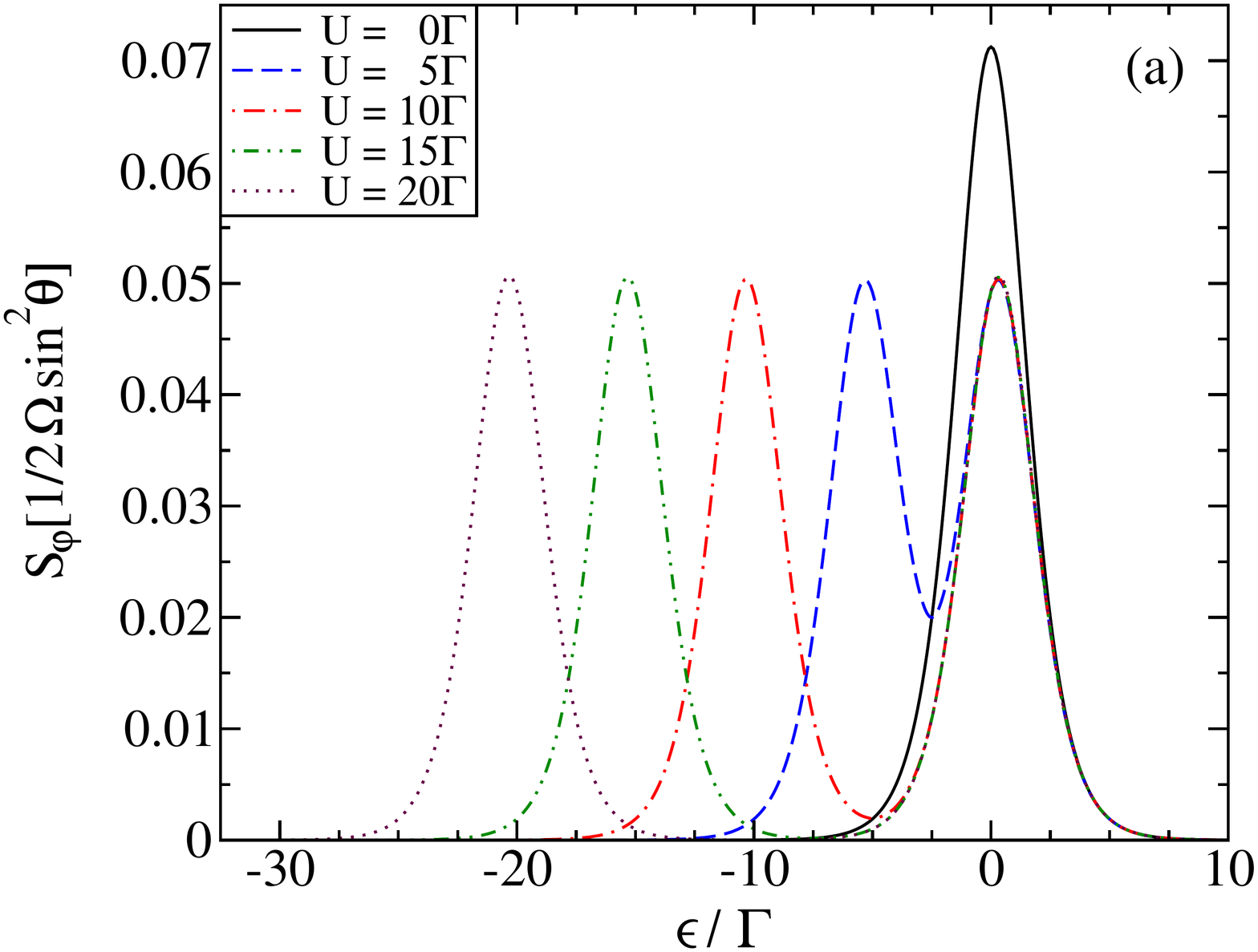} 
	\includegraphics[width=0.45\textwidth]{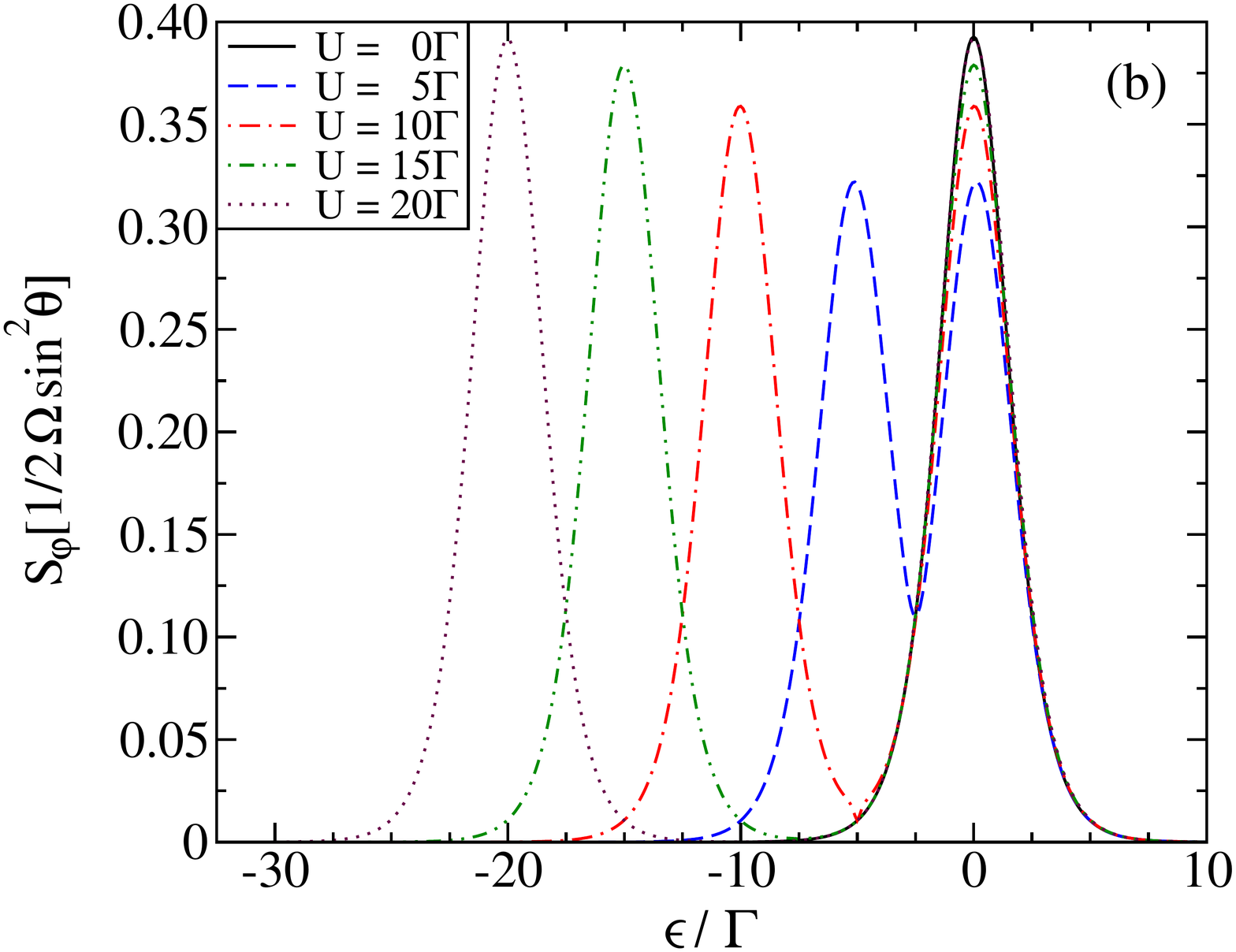} 
	\includegraphics[width=0.45\textwidth]{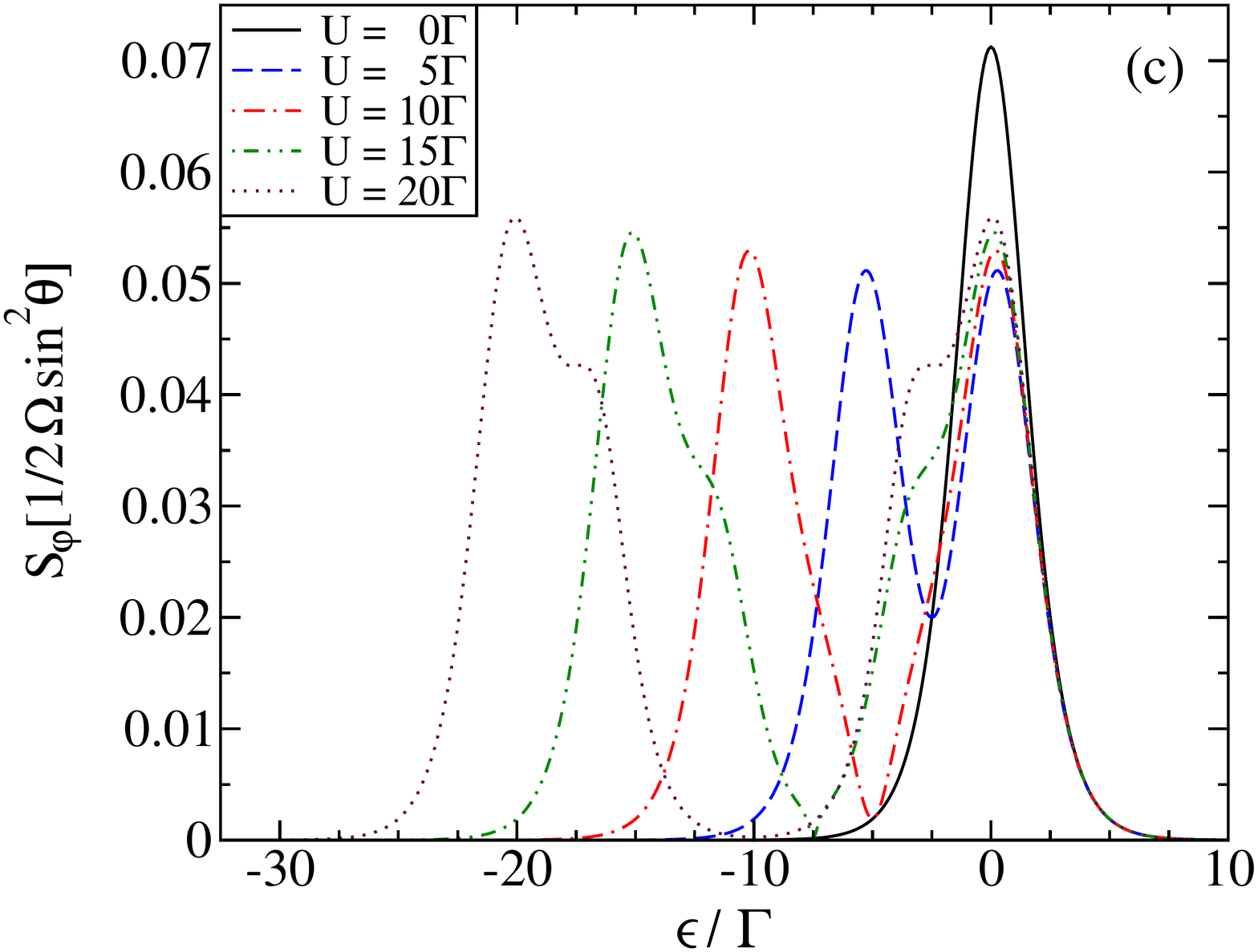} 
	\caption{(Color online) Pumped spin in units of $\frac{1}{2}  \Omega \sin^2\theta$ 	as a function of the dot  level position $\epsilon$ for different values of the Coloumb interaction $U$. 
	The temperature is chosen to be $k_\text{B}T= \Gamma$ and the ferromagnetic lead is assumed to be fully polarized ($p=1$).
	The different panels refer to the following cases: 
	(a) Weak coupling to the normal lead $(\Gamma_\text{F}=20\Gamma_\text{N})$;
	(b) Symmetric coupling to the leads $(\Gamma_\text{F}=\Gamma_\text{N})$;
	(c) Strong coupling to the normal lead $(20\Gamma_\text{F}=\Gamma_\text{N})$.  
	 Figure taken from Fig. 9 of Ref.~\cite{winkler_2013}.}
	\label{fig_spin_current}
\end{figure}

The pumped spin as a function of level position $\epsilon$ is shown in Fig.~\ref{fig_spin_current} for different values of the relative tunnel-coupling strength to the normal and ferromagnetic lead. We start in panel (a) with the case of  weak tunneling to the normal lead ($\Gamma_{\text{N}}\ll \Gamma_{\text{F}}$). In this situation the exchange field has a negligible effect ($A\approx 1$). The pumped spin has peaks located at $\epsilon=0$ and $\epsilon=-U$ (for $U=0$ only one resonance is present). 
The symmetric case ($\Gamma_{\text{N}}= \Gamma_{\text{F}}$) is shown in panel (b). The symmetric choice of the tunnel couplings  maximizes $G_0$ and hence the amplitude of the pumped spin.  
The increase in the peak amplitude as a function of interaction strength $U$ that is seen for $U>5 \Gamma$ reflects the behavior of the enhancement factor $A$. 
Finally, in panel (c) of Fig.~\ref{fig_spin_current} we show the case of strong coupling to the normal lead  $(\Gamma_{\text{N}}\gg \Gamma_{\text{F}})$, for which the exchange field strongly affects the pumped spin. 
In particular, the pumped spin is enhanced for values of the level position between the resonances and this leads to the appearances of  side peaks (shoulders), one below $\epsilon=0$ and one above $\epsilon=-U$.

\section{Spin pumping beyond the adiabatic regime\label{sec:NA}}
\label{section-na}
The adiabatically pumped spin is independent of the pumping frequency  $\Omega$ and hence in this regime the spin current  scales linearly with $\Omega$. 
Therefore, in order to achieve a sizable spin current, higher frequencies are desirable. 
Once $\Omega$ becomes of the order of the tunnel-coupling strength, pumping is beyond the adiabatic regime.
Moreover, non-adiabatic pumping introduces an additional control parameter, namely the pumping frequency, to steer charge and spin currents. 
In particular, it can be used to achieve pure spin pumping by tuning the pumping frequency such that the pumped charge current vanishes.

In the following, we study pumping through the same system as discussed in Sec.~\ref{sec:FDN}.
We again, restrict ourselves to the limit of weak tunneling, such that the lowest-order contribution in the tunnel-coupling strength is sufficient.
And again, we perform a systematic expansion in the pumping frequency. 
But this time, we do not only include the first two terms, instantaneous part and adiabatic correction.
Instead, we keep all orders in frequency.
This leads to a hierarchy for the kinetic equations, given by Eqs.~(\ref{rho_i small gamma}) for the instantaneous part and 
\begin{align}
	 \dot \varrho_t^{(k-1,-k+1)} + i [\bar H_{\text{dot}}, \varrho_t^{(k,-k)}] = W_t^{(0,1)} \varrho_t^{(k,-k)} \, .
\end{align}
The superscript on the left indicates the order in $\Omega$ and it is one for the adiabatic corrections and larger than one for the non-adiabatic ones.
We can solve for $\varrho_t^{(k,-k)}$ recursively starting from the instantaneous term $\varrho_t^{(0,-0)}$.  
The charge and spin currents can be expanded in powers of $\Omega$ as well.

In first order in the tunnel-coupling strengths, it turns out that it is possible to resum the expressions for the pumped charge and spin current to all orders in the adiabatic expansion.
As pumping parameters we choose $\{\Gamma_\text{N} , \varepsilon \}$.
Assuming weak pumping (bilinear response in $\Delta\Gamma_\text{N}$ and $\Delta\varepsilon$) we find that the pumped charge and spin take the form
\begin{align}
Q=Q^{\text{max}}\sin\left(\phi+\Delta\phi_{Q}\right) \\
S=S^{\text{max}}\sin\left(\phi+\Delta\phi_{S}\right) 
\end{align} 
where the amplitudes  ${Q}^{\text{max}}$ and $S^{\text{max}}$ are even functions of $\Omega$ while the phase shifts are odd functions of $\Omega$.
This implies that when expanding the pumped charge and spin in powers of $\Omega$, all
the even powers are proportional to $\sin \phi$ while the odd powers are proportional to $\cos \phi$. 
The adiabatic-pumping results are given by the zeroth-order contribution in $\Omega$ to the pumped charge and spin. 

\begin{figure}[htbp]
\begin{center}
\includegraphics[width=7.5cm]{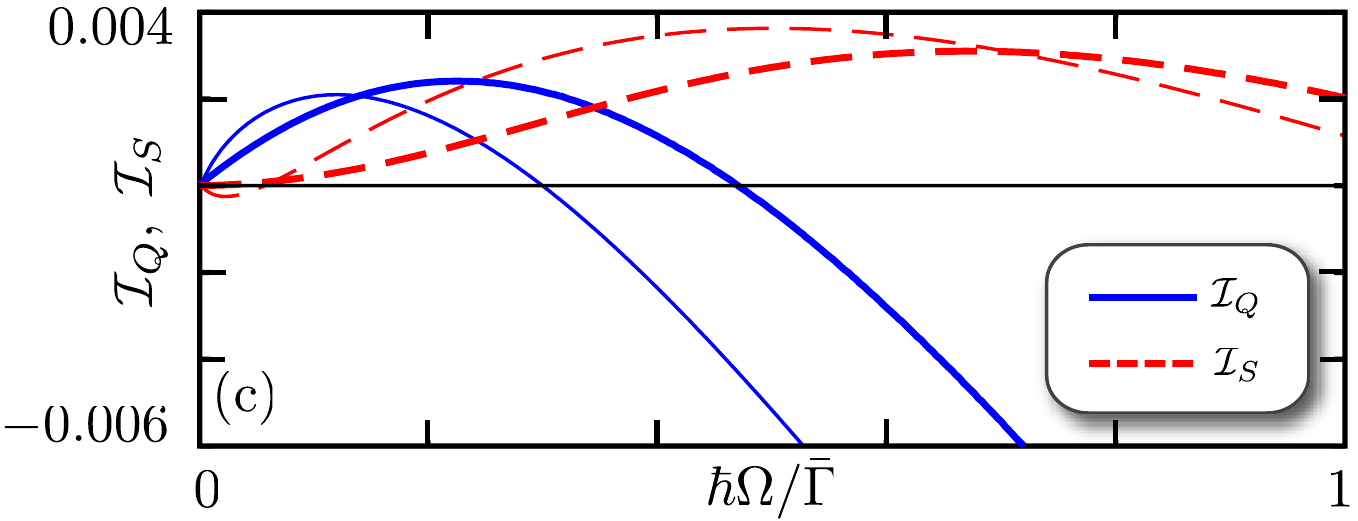}
\caption{(Color online) Average charge and spin currents $\mathcal{I}_{Q,S}$ as a function of
  $\hbar\Omega/\bar{\Gamma}$ for $p=0.4$ and pumping phase $\phi=\pi/9$. Thick lines refer to the weak-pumping regime and thin line to strong pumping (
 $\delta\Gamma_\text{N}=\delta\varepsilon=0.4\bar{\Gamma}$). Charge currents are in units of $\beta\delta\Gamma_{\text{N}}\delta\varepsilon/\hbar$ and spin currents 
  in units of $\beta\delta\Gamma_{\text{N}}\delta\varepsilon/2$. Other
  parameters: $\bar{\varepsilon}=k_B T\log{2}$, $U=9 \bar{\Gamma}$,
  $k_B T=3 \bar{\Gamma}$, and $\bar{\Gamma}_{\text{N}}/\bar{\Gamma}=1/2$.
  Figure taken from Fig. 2(c) of Ref.~\cite{cavaliere}.
  }\label{fig:nonad}
\end{center}
\end{figure}

An intriguing feature due to the non-adiabatic contributions to the pumped currents is the appearance of non-zero phase shifts $\Delta\phi_{Q(S)}$. Such phase shifts generically show up in nonadiabatically driven quantum systems, ranging from circuit QED \cite{wallraff}, optical lattices \cite{gommers}, spin-boson models \cite{grifoni}, and molecular systems \cite{astumian} to nanoelectromechanical devices \cite{pistolesi}. More importantly, the phase shifts for charge and spin differ from each other. This effect opens up the possibility  to generate a pure spin current, by tuning $\Omega$ such that $\phi+\Delta\phi_{Q}(\Omega)=0$ and consequently  $Q=0$. 
This is shown in Fig.~\ref{fig:nonad}, where the average charge and spin currents (indicated by a calligraphic font, i.e. $\mathcal{I}_{Q(S)}=Q(S) \Omega/(2\pi)$) are plotted as function of the pumping frequency $\Omega$. In particular, $\mathcal{I}_Q$ can be made to vanish for values of the frequency for which  $\mathcal{I}_S$ has a finite value. 

\section{Related issues}

The studies reported upon above have been extended in various directions. Here, we mention a few.

The properties of adiabatic pumping through a mesoscopic device depends on its spectral properties.
In the studies reported upon above, we have focused on quantum dots with large single-level spacing such that only one or two orbital levels participate in transport.
But also for the opposite limit of a continuous density of states, as realized in metallic single-electron transistors, our diagrammatic theory can be formulated and applied.
In the limit of weak tunnel coupling, the characteristics for the pumped charge current exhibits similar features as for the single-level quantum dot, with significant differences in situation in which the lowest-order pumping contribution is associated with a renormalization of the charge addition energy of the quantum dot~\cite{winkler_2009}.

An important issue in the context of adiabatic pumping is the question of how fast the system reacts to changes of the external system parameters.
Therefore, we studied the response of charge and spin in single-level quantum dots to sudden changes of the gate voltage in the limit of weak tunnel coupling. 
In the presence of Coulomb interaction, both the charge and the spin relaxation times depend on the dot's level position, but they differ from each other~\cite{relax-times_1}.
In addition, there is a third time scale associated with two-particle processes~\cite{relax-times_2}.

We have extended the diagrammatic formalism to include current-current correlations of adiabatic charge pumping. 
Performing a systematic perturbation expansion in the tunnel-coupling strength, we found Coulomb-interaction induced deviations from the fluctuation-dissipation theorem and discussed the influence of the asymmetry in the tunnel coupling on the noise properties~\cite{riwar}. 

When discussing spin pumping through quantum dots, the coherent superposition of spin-up and -down electrons may play an important role.
Similarly, coherence effects are an issue in Aharonov-Bohm interferometers that include quantum dots or in quantum dots coupled to superconducting leads. 
For both cases, we extended our diagrammatic technique and analyzed the pumping characteristics~\cite{hiltscher_ABI_pumping,hiltscher_CP_pumping}.

\begin{acknowledgement}
This topical review is based on work performed and published in fruitful collaborations with M. B\"uttiker, F. Cavaliere, L.D. Contreras-Pulido, R. Fazio, B. Hiltscher, A. Shnirman, J. Splettstoesser, and N. Winkler.
We, furthermore, acknowledge financial support from DFG via SPP 1285 (KO 1987/3), SFB 491, KO 1987/5, and the EU under grant No. 238345 (GEOMDISS).
\end{acknowledgement}

%

\begin{thebibliography}{[1]}

\bibitem{rev-zutic}I. \u{Z}uti\'c, J. Fabian, and S. Das Sarma, Rev. Mod. Phys. \textbf{76}, 323 (2004).

\bibitem{rev-awschalom} D.\,D. Awschalom and M.\,E. Flatte, Nat. Phys. \textbf{3}, 153 (2007).

\bibitem{rev-sinova} J. Sinova and I. \u{Z}ut\'c, Nat. Mater. \textbf{11}, 368 (2012).

\bibitem{brouwer}
  P.\,W.~Brouwer, Phys. Rev. B {\bf 58}, R10135 (1998).

\bibitem{buttiker-94} 
M. B\"uttiker, H. Thomas, and A. Pr\^etre, Z. Phys. B \textbf{94}, 133 (1994).

\bibitem{zhou-99}
  F.~Zhou, B.~Spivak, and B.~Altshuler, Phys. Rev. Lett. {\bf 82}, 608 (1999).

\bibitem{moskalets-01}
M. Moskalets, and M. B\"uttiker, Phys. Rev. B {\bf 64}, 201305 (2001).

\bibitem{makhlin-01}
  Yu. Makhlin and A.\,D.~Mirlin,
  Phys. Rev. Lett. {\bf 87}, 276803 (2001).

\bibitem{moskalets-02}
  M. Moskalets, and M. B\"uttiker, Phys. Rev. B {\bf 66}, 035306 (2002); Phys. Rev. B {\bf 66}, 205320 (2002).

\bibitem{entin-wohlman}
O. Entin-Wohlman, A. Aharony, and Y. Levinson, Phys. Rev. B {\bf 65}, 195411 (2002).

\bibitem{aleiner-98} 
  I.\,L. Aleiner and A.\,V. Andreev, Phys. Rev. Lett. \textbf{81}, 1286 (1998). 

\bibitem{aono-04} 
  T. Aono, Phys. Rev. Lett. \textbf{93}, 116601 (2004).

\bibitem{cota-94}
  E. Cota, R. Aguado, and G. Platero, Phys. Rev. Lett. {\bf 94}, 107202 (2005).

\bibitem{janineprl} 
  J. Splettstoesser, M. Governale, J. K\"onig, and R. Fazio, 
  Phys. Rev. Lett. \textbf{95}, 246803 (2005).
  
\bibitem{oreg} 
  E. Sela and Y. Oreg, Phys. Rev. Lett. {\bf 96}, 166802 (2006).

\bibitem{janineprb.06} 
  J. Splettstoesser, M. Governale, J. K\"onig, and R. Fazio, 
  Phys. Rev. B {\bf 74}, 085305 (2006).

\bibitem{silva} 
  D. Fioretto and A. Silva, Phys. Rev. Lett. {\bf 100}, 236803 (2008).

\bibitem{braun_burkard} 
  M. Braun and G. Burkard, Phys. Rev. Lett. {\bf 101}, 036802 (2008).

\bibitem{citro}
  F. Romeo, R. Citro, and M. Marinaro, Phys. Rev. B {\bf 78}, 245309 (2008). 

\bibitem{hernandez}
  A.\,R. Hernandez, F.\,A. Pinheiro, C.\,H. Lewenkopf, and E.\,R. Mucciolo, Phys. Rev. B {\bf 80}, 115311 (2009). 

\bibitem{yuge}
  T. Yuge, T. Sagawa, A. Sugita, and H. Hayakawa, Phys. Rev. B {\bf 86}, 235308 (2012). 

\bibitem{deus}
  F. Deus, F.\,R. Hernandez, and M.\,A. Continentino, J. Phys.: Condens. Matter {\bf 24}, 356001 (2012). 

\bibitem{switkes99} M. Switkes, C. M. Marcus, K. Campman, and A. C. Gossard, Science {\bf 283}, 1905 (1999).

\bibitem{giazotto} F. Giazotto, P. Spathis, S. Roddaro, S. Biswas, F. Taddei, M. Governale, L. Sorba,     Nature Physics \textbf{7}, 857 (2011).

\bibitem{pothier92} H. Pothier, P. Lafarge, C. Urbina, D. Esteve, and M. H. Devoret, Europhys. Lett. {\bf 17}, 249 (1992).

\bibitem{watson_experimental_2003}
  S.\,K. Watson, R.\,M. Potok, C.\,M. Marcus, and V. Umansky, Phys. Rev. Lett. {\bf 91}, 258301 (2003).

\bibitem{blumenthal07} M. D. Blumenthal, B. Kaestner, L. Li, S. Giblin, T. J. B. M. Janssen, M. Pepper, D. Anderson, G. Jones, and  D. A. Ritchie, Nature Phys. {\bf 3}, 343 (2007).

\bibitem{buitelaar08} M. R. Buitelaar, V. Kashcheyevs, P. J. Leek, V. I. Talyanskii, C. G. Smith, D. Anderson, G. A. C. Jones, J. Wei, and D. H. Cobden, Phys. Rev. Lett. {\bf 101}, 126803 (2008).

\bibitem{maisi09}V. F. Maisi, Yu. A. Pashkin, S. Kafanov, J. S. Tsai, and J. P. Pekola, New J. Phys. {\bf 11} 113057 (2009).

\bibitem{mirovsky10} P. Mirovsky, B. Kaestner, C. Leicht, A. C. Welker, T. Weimann, K. Pierz, H. W. Schumacher, Appl. Phys. Lett. {\bf 97}, 252104 (2010). 	

\bibitem{riwar2010}
  R.-P. Riwar and J. Splettstoesser, Phys. Rev. B {\bf 82}, 205308 (2010).

\bibitem{reckermann}
  F. Reckermann, J. Splettstoesser, and M.\,R. Wegewijs, Phys. Rev. Lett. {\bf 104}, 226803 (2010).

\bibitem{calvo}
  H.\,L. Calvo, L. Classen, J. Splettstoesser, and M.\,R. Wegewijs, Phys. Rev. B {\bf 86}, 245308 (2012).

\bibitem{mucciolo_2002}
  E.\,R. Mucciolo, C. Chamon, and C.\,M. Marcus, Phys. Rev. Lett. {\bf 89}, 146802 (2002).

\bibitem{fransson_2010}
  J. Fransson and M. Galperin, Phys. Rev. B {\bf 81}, 075311 (2010).

\bibitem{fransson_2013}
  B.\,O. Jahn, H. Ottosson, M. Galperin, and J. Fransson, ACS Nano {\bf 7}, 1064 (2013).

\bibitem{hamaya}
  K. Hamaya, S. Masubuchi, M. Kawamura, T. Machida, M. Jung, K. Shibata, K. Hirakawa, T. Taniyama, S. Ishida, and Y. Arakawa, Appl. Phys. Lett. {\bf 90}, 053108 (2007).
  
\bibitem{deshmukh_2002}
  M.\,M. Deshmukh and D.\,C. Ralph, Phys. Rev. Lett. {\bf 89}, 266803 (2002).
  
\bibitem{mitani_2008}  
   S. Mitani, Y. Nogi, H. Wang, K. Yakushiji, F. Ernult, and K. Takanashi, Appl. Phys. Lett. {\bf 92}, 152509 (2008).
   
\bibitem{fert_2009}
  A. Bernand-Mantel, P. Seneor, K. Bouzehouane, S. Fusil, C. Deranlot, F. Petroff, and A. Fert, Nat. Phys. {\bf 5}, 920 (2009).
  
\bibitem{davidovic_2010}
  F.\,T. Birk and D. Davidovic, Phys. Rev. B {\bf 81}, 241402 (2010).
  
\bibitem{hofstetter_2010}
  L. Hofstetter, A. Geresdi, M. Aegesen, J. Nygard, C. Sch\"onenberger, and S. Csonka, Phys. Rev. Lett. {\bf 104}, 246804 (2010).
 
\bibitem{schoenenberger_2005}
  S. Sahoo, T. Kontos, J. Furer, C. Hoffmann, M. Gr\"aber, A. Cottet, and C. Sch\"onenberger, Nat. Phys. {\bf 1}, 99 (2005).

\bibitem{lindelof_2008}
  J.\,R. Hauptmann, J. Paaske, and P.\,E. Lindelof, Nat. Phys. {\bf 4}, 373 (2008).

\bibitem{pasupathy_2004}
  A.\,N. Pasupathy, R.\,C. Bialczak, J. Martinek, J.\,E. Grose, L.\,A.\,K. Donev, P.\,L. McEuen, and D.\,C. Ralph, Science {\bf 306}, 86 (2004).

\bibitem{Nowack_tune_SOI_exp} 
  K.\,C. Nowack, F.\,H.\,L. Koppens, Y.\,V. Nazarov, and L.\,M.\,K. Vandersypen, 
  Science \textbf{318}, 1430 (2007).
  
 \bibitem{Fasth_SO_in_QD_exp} 
  C. Fasth, A. Fuhrer, L. Samuelson, V.\,N. Golovach, and D. Loss, 
  Phys. Rev. Lett. \textbf{98}, 266801 (2007).

 \bibitem{Nilsson_SO_in_QD_exp}
  H.\,A. Nilsson, P. Caroff, C. Thelander, M. Larsson, J.\,B. Wagner, L. Wernersson, L. Samuelson, and H.\,Q. Xu, 
  Nano Lett. \textbf{9}, 3151 (2009).

  \bibitem{Katsaros_SO_in_QD_exp}
  G. Katsaros, P. Spathis, M. Stoffel, F. Fournel, M. Mongillo, V. Bouchiat, F. Lefloch, A. Rastelli, O.\,G. Schmidt, and S. De Franceschi, 
  Nat. Nano. \textbf{5}, 458 (2010).
  
\bibitem{Takahashi_SO_in_QD_exp} 
 S. Takahashi, R.\,S. Deacon, K. Yoshida, A. Oiwa, K. Shibata, Y. Tokura, and S. Tarucha,
 Phys. Rev. Lett. \textbf{104}, 246801 (2010).
 
 \bibitem{Lai_SO_in_QD} 
 R.\,A. Lai, H.\,O.\,H. Churchill, and C.\,M. Marcus, 
Nature Nanotech. \textbf{8}, 565 (2013).

\bibitem{splettstoesser_2008}
    J. Splettstoesser, M. Governale, and J. K\"onig, Phys. Rev. B \textbf{77}, 195320 (2008).

\bibitem{winkler_2013}
    N. Winkler, M. Governale, and J. K\"onig, Phys. Rev. B \textbf{87}, 155428 (2013).
  
  \bibitem{Sharma_SOI_spinpumping} 
  P. Sharma and P.\,W. Brouwer, 
  Phys. Rev. Lett. \textbf{91}, 166801 (2003).
 
    \bibitem{Governale_SOI_spinpumping}
  M. Governale, F. Taddei, and R. Fazio, 
  Phys. Rev. B \textbf{68}, 155324 (2003).
  
\bibitem{brosco_prediction_2010}%
     V. Brosco, M. Jerger, P. San-Jos\'{e}, G. Zar\'{a}nd,  A. Shnirman, and  G. Sch\"on, Phys.
     Rev. B \textbf{82}, 041309 (2010).

  \bibitem{Eto_spin_filter} 
  M. Eto and T. Yokoyama, 
  J. Phys. Soc. Jpn. \textbf{79}, 123711 (2010).

\bibitem{grap_interplay_2012}%
  S. Grap, V. Meden, and S. Andergassen, Phys. Rev. B \textbf{86}, 035143 (2012).

\bibitem{droste_josephson_2012}%
   S. Droste, S. Andergassen, and J. Splettstoesser, J. Phys.: Condens. Matter \textbf{24}, 415301 (2012).

\bibitem{rojek} 
  S. Rojek, J. K\"onig, and A. Shirman, Phys. Rev. B {\bf 87}, 075305 (2013).

\bibitem{avron_geometry_2000}%
    J.\,E. Avron, A. Elgart, G.\,M. Graf, and L. Sadun, Phys. Rev. B \textbf{62}, R10618 (2000).

\bibitem{schoeller_mesoscopic_1994}%
  H. Schoeller and G. Sch\"on, Phys. Rev. B \textbf{50}, 18436 (1994).

 \bibitem{knig_resonant_1996}%
    J. K\"onig,  J. Schmid, H. Schoeller, and G. Sch\"on, Phys. Rev. B \textbf{54}, 16820 (1996).

 \bibitem {knig_zero-bias_1996}%
   J. K\"onig, H. Schoeller, and G. Sch\"on, Phys. Rev. Lett. \textbf{76},  1715 (1996).
   
\bibitem{moca_2013}
  C.\,P. Moca, A. Alex, A. Shnirman, and G. Zar\'{a}nd, arXiv:1307.3416.

\bibitem{tserkovnyak_enhanced_2002}
  Y. Tserkovnyak, A. Brataas, and G.\,E.\,W. Bauer, Phys. Rev. Lett. {\bf 88}, 117601 (2002).
  
\bibitem{brataas_spin_2002}
  A. Brataas, Y. Tserkovnyak, G.\,E.\,W. Bauer, and B.\,I. Halperin, Phys. Rev. B {\bf 66}, 060404 (2002).
  
\bibitem{tserkovnyak_spin_2002}
  Y. Tserkovnyak, A. Brataas, and G.\,E.\,W. Bauer, Phys. Rev. B {\bf 66}, 224403 (2002).
  
\bibitem{mahfouzi_microwave_2010}
  F. Mahfouzi, B.\,K. Nikolic, S.\,H. Chen, and C.\,R. Chang, Phys. Rev. B {\bf 82}, 195440 (2010).
 
\bibitem{costache_electrical_2006}
  M.\,V. Costache, M. Sladkov, S.\,M. Watts, C.\,H. van der Wal, and B.\,J. van Wees, Phys. Rev. Lett. {\bf 97}, 216603 (2006).

\bibitem{bender_tserkovnyak_brataas_2010}
  S.\,A. Bender, Y. Tserkovnyak, and A. Brataas, Phys. Rev. B {\bf 82}, 180403(R) (2010).

\bibitem{koenig_interaction_2003}
  J. K\"onig and J. Martinek, Phys. Rev. Lett. {\bf 90}, 166602 (2003).
  
\bibitem{martinek_kondo_qd_2003}
  J. Martinek, Y. Utsumi, H. Imamura, J. Barnas, S. Maekawa, J. K\"onig, and G. Sch\"on, Phys. Rev. Lett. {\bf 91}, 127203 (2003). 
  
\bibitem{braun_theory_2004}  
   M. Braun, J. K\"onig, and J. Martinek, Phys. Rev. {\bf B} 70, 195345 (2004).

\bibitem{cavaliere} 
  F. Cavaliere, M. Governale, and J. K\"onig, Phys. Rev. Lett. {\bf 103}, 136801 (2009).

\bibitem{wallraff} 
 A. Wallraff, D.\,I. Schuster, A. Blais, L. Frunzio, R.-S. Huang, J. Majer, S. Kumar, S.\,M. Girvin, and R.\,J. Schoelkopf, Nature \textbf{431}, 162 (2004). 

\bibitem{gommers}
  R. Gommers, S. Bergamini, and F. Renzoni, Phys. Rev. Lett. {\bf 95}, 073003 (2005).
  
\bibitem{grifoni}
  M. Grifoni and P. H\"anggi, Phys. Rev. Lett. {\bf 76}, 1611 (1996).    

\bibitem{astumian} 
 R.\,D. Astumian and I. Derenyi, Phys. Rev. Lett. \textbf{86}, 3859 (2001).

\bibitem{pistolesi} 
 F. Pistolesi and R. Fazio, Phys. Rev. Lett. \textbf{94}, 036806 (2005). 

\bibitem{winkler_2009}
  N. Winkler, M. Governale, and J. K\"onig, Phys. Rev. B {\bf 79}, 235309 (2009).

\bibitem{relax-times_1} 
  J. Splettstoesser, M. Governale, J. K\"onig, and M. B\"uttiker,
  Phys. Rev. B {\bf 81}, 165318 (2010).

\bibitem{relax-times_2} 
  L\,D. Contreras-Pulido, J. Splettstoesser, M. Governale, J. K\"onig, and M. B\"uttiker,
  Phys. Rev. B {\bf 85}, 075301 (2012).

\bibitem{riwar} 
  R.\,P. Riwar, J. Splettstoesser, and J. K\"onig, 
  Phys. Rev. B {\bf 87}, 195407 (2013).
  
\bibitem{hiltscher_ABI_pumping} 
  B. Hiltscher, M. Governale, and J. K\"onig, 
  Phys. Rev. B \textbf{81}, 085302 (2010).

\bibitem{hiltscher_CP_pumping} 
  B. Hiltscher, M. Governale, J. Splettstoesser, and J. K\"onig, 
  Phys. Rev. B \textbf{84}, 155403 (2011).
  
\end{thebibliography}
%

\end{document}